\documentclass[11pt]{article}
\usepackage{color,amsmath,amssymb,graphicx}
\usepackage{cancel}

\definecolor{refkey}{rgb}{1,0.5,0.5}
\definecolor{labelkey}{rgb}{0.5,1,0.5}
 \numberwithin{equation}{section}
\date{\today}

\usepackage{epstopdf}
\newcommand{\nc}{\newcommand}
\nc{\nit}{\noindent}
\nc{\nn}{\nonumber}
\newcommand{\be}{\begin{equation}}
\newcommand{\ba}{\begin{eqnarray}}
\newcommand{\ea}{\end{eqnarray}}
\newcommand{\ee}{\end{equation}}

\newcommand{\sech}{{\rm sech}}
\nc{\D}{\partial}
\nc{\diff}[2]{\frac{d #1}{d #2}}
\nc{\diffn}[3]{\frac{d^{#3} #1}{d {#2}^{#3}}} 
\nc{\pdiff}[2]{\frac{\partial #1}{\partial #2}} 
\nc{\pdiffn}[3]{\frac{\partial^{#3} #1}{\partial{#2}^{#3}}} 
\nc{\abs}[1] {\lvert #1 \rvert} 

\nc{\st}{{\star}}
\nc{\cE}{{\cal E}} 
\nc{\cF}{{\cal F}} 
\nc{\cH}{{\cal H}}
\nc{\cN}{{\cal N}} 
\nc{\cV}{{\cal V}} 
\nc{\cQ}{{\cal Q}} 
\nc{\cGin}{{\cal G}_{\rm in}} 
\nc{\cGout}{{\cal G}_{\rm out}} 
\nc{\cO}{{\cal O}}
\nc{\Lav}{{\cal L}_{\rm av}}
\nc{\cL}{{\cal L}}
\nc{\cB}{{\cal B}}
\nc{\cZ}{{\cal Z}} 
\nc{\cT}{{\cal T}} 
\nc{\cY}{{\cal Y}} 
\nc{\vD}{{\vec D}} 
\nc{\vE}{{\vec E}} 
\nc{\vB}{{\vec B}} 
\nc{\vH}{{\vec H}} 
\nc{\ty}{{\tilde y}} 
\nc{\psit}{{ \tilde{\psi} } } 
\nc{\tpsi}{{\Phi}} 
\nc{\tu}{{\tilde u}} 
\nc{\tV}{{\tilde V}} 
\nc{\tVhat}{\hat{\tilde V}} 
\nc{\bx}{{\bf x}} 
\nc{\bk}{{\bf k}} 
\nc{\bX}{{\bf X}} 
\nc{\bXYZ}{{\bf XYZ}} 
\nc{\bY}{{\bf Y}} 
\nc{\bZ}{{\bf Z}} 
\nc{\bF}{{\bf F}} 
\nc{\bS}{{\bf S}} 
\nc{\dV}{{\delta V}} 
\nc{\dVN}{{\delta V_N}} 
\nc{\dVNv}{{\delta \check{V}_N}} 
\nc{\dVt}{{\delta\tilde{V}}} 
\nc{\dv}{{\delta v}} 
\nc{\dE}{{\delta E}} 
\nc{\dPsi}{{\delta\Psi}} 
\nc{\pp}{\perp} 
\nc{\order}{{\cal O}}
\nc{\Eplus}{E_+} 
\nc{\Eminus}{E_-} 
\nc{\Epm}{E_\pm}
\nc{\Vav}{V_{\rm av}}
\nc{\Rin}{R_{\rm in}}
\nc{\Rout}{R_{\rm out}}
\nc{\eplus}{e_+} 
\nc{\eminus}{e_-} 
\nc{\epm}{e_\pm}
\nc{\eps}{\epsilon} 
\nc{\half}{{1\over2}} 
\nc{\veps}{\varepsilon} 
\nc{\vnabla}{{\vec\nabla}} 
\nc{\G}{\Gamma} 
\nc{\sphi}{\Phi} 
\nc{\w}{\omega} 
\nc{\mh}{h} 
\nc{\mg}{g} 
\nc{\sgn}{{\rm sgn}} 
\nc{\vphi}{\varphi}
\nc{\tlambda}{\tilde\lambda} 

\renewcommand{\d}{\delta}

\nc{\g}{\gamma} 
\nc{\ol}{\overline}

\newtheorem{theo}{Theorem}[section]

\newtheorem{lem}{Lemma}[section]

\newtheorem{remark}{Remark}[section]
\def\R{\mathbb{R}}

\nc{\pT}{\partial_T} 
\nc{\pz}{\partial_z}
\nc{\pt}{\partial_t} 
\nc{\la}{\langle} 
\nc{\ra}{\rangle} 
\nc{\infint}{\int_{-\infty}^{\infty}}
\nc{\halfwidth}{6.5cm}
\nc{\figwidth}{10cm}
\nc{\essspec}{\sigma_{\rm ess}}
\nc{\sqrtE}{\mu}
\nc{\resolv}{R}
\nc{\proofend}{$\square$}
\nc{\localize}{\chi}
\nc{\localizehat}{\hat{\chi}}
\nc{\localizeinv}{\chi^{-1}}
\nc{\localizet}{\tilde{\chi}}
\nc{\localizethat}{\hat{\localizet}}
\nc{\localizetinv}{\tilde{\chi}^{-1}}
\nc{\mloc}{M}
\nc{\mtloc}{\tilde{M}}
\nc{\smoothinv}{\langle D \rangle}
\nc{\smoothinvhat}[1][\xi]{\langle #1 \rangle}
\nc{\smooth}{\langle D \rangle^{-1}}
\nc{\smoothhat}[1][\xi]{\langle #1 \rangle^{-1}}
\nc{\TR}{T_R}
\nc{\TRl}{T_{R,l}}
\nc{\TRE}{T_R}
\nc{\TRlE}{T_{R,l}}
\nc{\dTRE}{\delta\TRE}
\nc{\TV}[1][V]{T_{#1}}
\nc{\SR}{S_R}
\nc{\SV}[1][V]{S_{#1}}
\nc{\twopin}{(2\pi)^n}
\nc{\twopiminusn}{(2\pi)^{-n}}
\nc{\msmooth}{m}
\nc{\resD}{\mathcal{L}}
\nc{\bessker}{G}
\nc{\remker}{\mathcal{R}}
\nc{\TRI}{A_{\rm I}}
\nc{\TRII}{A_{\rm II}}
\nc{\TRIIa}{A_{\rm II}^{(a)}}
\nc{\TRIIb}{A_{\rm II}^{(b)}}
\nc{\TRIII}{A_{\rm III}}
\nc{\genBessel}{\mathcal{J}}
\nc{\bddcomm}{\mathcal{C}}
\nc{\Fhat}{\hat{F}}
\nc{\fhat}{\hat{f}}
\nc{\ghat}{\hat{g}}
\nc{\FM}{F_2^{(1)}}
\nc{\FR}{F_2^{(2)}}
\nc{\xiunder}{\underline{\xi}}
\nc{\xiover}{\overline{\xi}}
\nc{\Psizerobar}{\overline{\Psi}_0}
\nc{\psizerobar}{\overline{\psi}_0}
\nc{\Vt}{\tilde{V}}
\nc{\Psit}{\tilde{\Psi}}
\nc{\Et}{\tilde{E}}
\nc{\dEcoef}{C_{dE}}
\nc{\Etwohomog}{E_2^{\rm(homog)}}
\nc{\dEone}{\dE^{(1)}}
\nc{\dEtwo}{\dE^{(2)}}
\nc{\dEtwomain}{N_2^{\rm{(PT,main)}}}
\nc{\dEtworem}{N_2^{\rm{(PT,rem)}}}
\nc{\fbar}{\overline{f}}
\nc{\leftfn}{f^{(L)}}
\nc{\rightfn}{f^{(R)}}
\nc{\partialwaveG}{G}
\nc{\supp}{\operatorname{supp}}
\nc{\rt}{\tilde{r}}

\nc{\PU}{\chi}
\nc{\gt}{\tilde{g}}
\nc{\indicator}{\boldsymbol{1}}
\nc{\Ai}{\mathop{\mathrm{Ai}}\nolimits}
\nc{\thetat}{\tilde{\theta}}

\numberwithin{equation}{section}

\date{\today}
\begin{document}
\title{Localized States and Dynamics in the\\ Nonlinear Schr\"odinger / Gross-Pitaevskii Equation
}
\author{ M.I.~Weinstein
\thanks{This work is supported in part by U.S. NSF grant DMS-10-08855 and
DMS-14-12560.}\\
Department of Applied Physics and Applied Mathematics,\\
and Department of Mathematics\\
Columbia University\\
New York, NY 10027}

\baselineskip=18pt
\maketitle
%
%
\thispagestyle{empty}
%

\section{Introduction}\label{sec:introduction}


 {\it Nonlinear dispersive waves} are wave phenomena resulting from the interacting effects of nonlinearity and dispersion. {\it Dispersion} refers to the property that waves of different wavelengths travel at different velocities. This property may, for example, be due to the material properties of the medium, {\it e.g.} chromatic dispersion \cite{Agrawal:95},  or to the geometric arrangement of material constituents, {\it e.g.} a periodic medium with Floquet-Bloch {\it band dispersion} \cite{Ashcroft-Mermin:76}. Nonlinearity distorts the shape of a localized structure or creates amplitude dependent non-uniformities in phase. This may  concentrate or localize energy in a region of space (attractive / focusing nonlinearity) or tend to expel energy from compact sets (repulsive / defocusing nonlinearity). In electromagnetics, nonlinearities may arise due to the intensity dependence of dielectric parameters ({\it e.g.} Kerr effect \cite{Agrawal:95}).  Physical phenomena in which the effects of both dispersion and nonlinearity play a role are ubiquitous. Some examples are: (a) long waves of small amplitude at a water-air interface \cite{Whitham:74}, (b) a nearly mono-chromatic laser beam propagating through air, glass or water \cite{Fibich:09}, (c) light-pulses propagating through optical fiber waveguides \cite{Agrawal:95} and (d) the macroscopic dynamics of weakly correlated quantum particles in a Bose-Einstein condensate; see, for example,
 \cite{PS:03,ESY:07,Grillakis-Machedon-Margetis:10}. Interest in nonlinear dispersive waves and their interaction with nonhomogeneous media ranges from Fundamental to Applied Science with great promise in engineering / technological applications due to major advances in materials science and micro- and nano-structure fabrication techniques; see, for example, \cite{Weinstein:07}.

 An important feature of many nonlinear dispersive systems is the prevalence of coherent structures.   They range from phenomena we can only passively observe in nature, such as  large-scale localized meteorological events, to those we can control, such as the flow of energy in computer chips and other nano-patterned media.  In general, a coherent structure is one which has a spatially localized core, {\it e.g.} a solitary wave concentrated in a compact region of space or a front-like state which transitions between different asymptotic states across a spatially compact transition region.
 %

 We focus on a key model in the mathematical theory of nonlinear dispersive waves, the
 {\it nonlinear Schr\"odinger / Gross Pitaevskii} (NLS/GP) equation:

 \begin{align}
  i\D_t \Phi &=-\Delta \Phi\  +\ V(x)\Phi +  g|\Phi|^2\Phi\ .
 \label{nls-gp}\end{align}
 In particular, we consider  its {\it nonlinear bound states}.
 Here, $\Phi(x,t)$ is a complex-valued function for $x\in\R^d$ and $t\in\R$. $V(x)$ is a real-valued ``potential'',
 and $g$ is a real parameter.
 The following table lists the physical significance of $\Phi$, $V$ and $g$ in the settings of nonlinear optics and macroscopic quantum physics:

 \begin{center}
\sf{\begin{tabular}{|l |  r  | l|  r l|}
\hline
{\bfseries } & {\bfseries  Nonlinear Optics} &   {\bfseries Macroscopic Quantum Physics}\\
\hline
$\Phi(x,t)$ ~ & Electric field complex envelope & Macroscopic quantum wave function \\%
 $V(x)$ & Refractive index & Magnetic trap  \\
$g$ &  Kerr nonlinearity coefficient, $g<0$ & Microscopic 2-body scattering length ($g>0$ or $g<0$)\\
\hline
\end{tabular}
}
\end{center}

\subsection{Outline }
Basic mathematical properties of NLS and NLS/GP are discussed in section \ref{nls-nlsgp}. In section \ref{bound-states} nonlinear bound states and aspects of their stability theory are discussed from variational and bifurcation perspectives. Examples are also presented and the particular cases of $V(x)$ given by a single-well, double-well potential and periodic potential are discussed in detail.  In section \ref{soliton-defect-intrxns} we discuss  particle-like dynamics of solitons interacting with a potential over a large, but finite, time interval. In sections \ref{res-rad-damp} and \ref{gss-ep} we consider the very long time behavior of solutions to NLS / GP. In particular, the focus is placed on the important resonant radiation damping mechanism that drives the relaxation of the system to a nonlinear  ground state and underlies the phenomena of {\it Ground State Selection} and {\it Energy Equipartition}. Sections \ref{simple-1}-\ref{simple-2} discuss linear and nonlinear {\it toy minimal models}, which illustrate these mechanisms. Regarding the overall style of this article,  we seek to emphasize the key ideas, and therefore do not present all detailed technical hypotheses, leaving that to the references.
\medskip

{\bf Acknowledgements:} The author would like to thank J. Marzuola and E. Shlizerman for stimulating discussions and helpful comments on this article. He also wishes to thank Bjorn Sandstede and the referees for their careful reading and suggested improvements.

\section{NLS and NLS/GP}\label{nls-nlsgp}

In this section we review basic properties of NLS/GP:
\begin{equation}
i\D_t\Phi = -\Delta\Phi\ + V(x)\Phi\ + \ g|\Phi|^{2\sigma}\Phi \ .
\label{NLS-GP}\end{equation}
Here, $g$ is a real parameter and we  assume that $V$ is a bounded and smooth real-valued potential.
 The equation \eqref{NLS-GP} is often called the  nonlinear Schr\"odinger / Gross-Pitaevskii (NLS/GP).

We consider solutions to the initial value problem for \eqref{NLS-GP} with initial conditions
\begin{equation}
\Phi(x,0) = \Phi_0(x) \in H^1(\R^d).
\label{Phi-data}\end{equation}
The basic well-posedness theorem states that for $d=1,2$ and all $0\le\sigma<\infty$ and for $d\ge3$ and $0\le\sigma<2/(d-2)$, the
 initial value problem \eqref{NLS-GP}-\eqref{Phi-data}  has a unique local-in-time solution $\Phi\in C^0([0,T);H^1(\R^d))$ \cite{Kato:87,Sulem-Sulem:99,Bourgain:99,Tao:06,Fibich:14}.

NLS / GP is a Hamiltonian system, expressible in the form
\begin{equation}
i\D_t\Phi = \frac{\delta \cH[\Phi,\Phi^*]}{\delta \Phi^*}\ ,
\label{NLS-GP-ham}
\end{equation}
with Hamiltonian energy functional
\begin{equation}\label{cH-def}
\cH[F,F^*]= \int_{\R^d} \nabla F(x)\cdot \nabla F^*(x) + V(x)\ F(x) F^*(x)\ +\  \frac{g}{\sigma+1} \left(F(x)F^*(x)\right)^{\sigma+1}\ dx.
\end{equation}
The solution of the initial value problem satisfies the conservation laws:
\begin{align}
\cH[\Phi(\cdot,t)]\ &=\ \cH[\Phi_0]\label{conserved-H}\\
\cN[\Phi(\cdot,t)]\ &=\ \cN[\Phi_0]\label{conserved-N}
\end{align}
where $\cH[\cdot]$ is defined in \eqref{cH-def} and
\begin{equation}
\cN[F,F^*]\ =\ \int_{\R^d}\ F(x) F^*(x)\ dx\ .
\label{Ndef}
\end{equation}
The conserved integrals $\cH$ and $\cN$ are associated, respectively, with the invariances
$t\mapsto t+t_0,\ t_0\in\R$ (time-translation) and $\Phi\mapsto\Phi e^{i\theta},\ \theta\in\R$.

For the nonlinear Schr\"odinger equation (NLS), where $V\equiv0$, we have the additional symmetries
\begin{enumerate}
\item $\Phi(x,t)\mapsto \Phi(x+x_0,t),\ x_0\in\R^d$\ (Translation invariance)
\item $\Phi(x,t)\mapsto \mathcal{G}_\xi\left[\Phi\right](x,t)=e^{i\xi(x-\xi t)}\ \Phi(x-2\xi t,t),\ \xi\in\R^d$\ (Galilean invariance)
\end{enumerate}
For NLS with the homogeneous power nonlinearity in \eqref{NLS-GP} the equation is also dilation invariant:
\begin{equation}
\Phi(x,t)\mapsto \lambda^{1\over\sigma}\Phi(\lambda x,\lambda^2 t)
\label{dilation}\end{equation}

 The case where $g<0$ is called the {\it focusing case} and the case where $g>0$ is called the  {\it defocusing case}. This terminology refers to the nonlinear term's tendency to focus (localize) energy or defocus (spread) energy. Independently of the nonlinear term, the linear potential
 may be an attractive potential, which concentrates energy, or a repulsive potential which allows energy to spread.

Concerning global-in-time behavior of solutions, the initial value problem is globally well-posed for any $H^1$ initial condition if either $g>0$ or if $g<0$ and $\sigma<2/d$ (subcritical case).  For $g<0$ and $\sigma\ge2/d$ solutions with $H^1$ initial data may develop singularities in finite time (blow up). See, for example, \cite{Kato:87,Sulem-Sulem:99,Bourgain:99,Tao:06,Weinstein:83,Fibich:14}.

\section{ Bound States\ -\ Linear and Nonlinear}\label{bound-states}

\subsection{Linear bound states}

We start with  the linear Schr\"odinger equation:
\begin{align}
i\D_t\Psi\ &= \left( -\Delta + V(x) \right)\Psi \ = H\ \Psi\label{linear-schr}
\end{align}
Here $V$ denotes a smooth and real-valued potential satisfying

 \begin{equation}
V(x)\to0\ \textrm{sufficiently rapidly as}\ |x|\to\infty\ .
\label{Vdecays}
\end{equation}
Bound states are solutions of the form $\Psi = e^{-iEt}\psi(x)$, where $\psi$ satisfies the eigenvalue problem
\begin{equation}
 H\psi\ =\ (-\Delta + V)\psi\ =\  E\psi ,\ \ \psi\in H^1(\R^d)\ .
 \label{lin-evp}
 \end{equation}

We expect the eigenvalue problem \eqref{lin-evp} to have a non-trivial solutions if $V$ is an attractive
``potential well''; intuitively, the set where $V(x)\le0$ should be sufficiently wide and deep.

The following result gives a condition on the existence of a ground state, a nontrivial solution of \eqref{lin-evp} of minimial energy, $E$
\cite{Courant-Hilbert-I:53,Lieb-Loss:97}\ .
\begin{theo}\label{linear-gs}
Denote by
\begin{equation}
E_{0\star}\ =\ \inf_{\int_{\R^d} |f|^2 =1} \langle Hf,f\rangle \equiv \inf_{\int_{\R^d} |f|^2 =1}
 \int_{\R^d}\ \left|\nabla f(x)\right|^2 + V(x)|f(x)|^2\ dx.
 \nn\end{equation}
 If  $-\infty<E_{0\star}<0$
  then  the infimum is attained at a positive function $\psi_{0\star}\in H^1(\R^d)$,
 which is a solution of
\begin{equation}
 H\psi_{0\star}=E_{0\star}\psi_{0\star},\ \  E_{0\star}=\left\langle H\psi_{0\star},\psi_{0\star}\right\rangle
 \nn\end{equation}
 \end{theo}

\subsection{Nonlinear bound states}

Here we consider spatially localized solutions of  NLS/GP of the form
\begin{equation}
\Phi(x,t)\ =\ e^{-iEt}\ \psi_E(x)\ ,
\label{sw-ansatz}
\end{equation}
where $\psi_E$ satisfies the nonlinear elliptic equation
\begin{equation}
(-\Delta +V)\psi_E + g|\psi_E|^{2\sigma}\psi_E\ =\ E\psi_E,\ \ \psi_E\in H^1(\R^d)
\label{psiE-eqn}
\end{equation}

Solutions of the nonlinear eigenvalue problem \eqref{psiE-eqn} are often called {\it nonlinear bound states} or  solitary waves. A positive and decaying solution is often called a nonlinear ground state.
 In the case where $V(x)$ is non-trivial, such states are often called nonlinear defect modes. In this case, the spatially varying potential, $V(x)$, is viewed as a localized defect in a homogeneous (translation-invariant) background.

The question of existence of nonlinear bound states of NLS / GP has been studied extensively by variational and bifurcation methods. See, for example,  \cite{Strauss:79,Berestycki-Lions:83,Weinstein:83,Lions:84,RW:88} as well as the discussion in sections \ref{double-well} \cite{KKSW:08} and \ref{spec-edge} \cite{Ilan-Weinstein:10}.

The following result, which we shall refer to later in this article, concerns bifurcation of a nonlinear ground state from the ground eigenstate of the linear operator  $-\Delta+V$; see, for example, \cite{RW:88,PilletWayne:97}

\begin{theo}\label{RWthm}
Assume $V$ is real-valued, smooth and rapidly decreasing at infinity and  that $-\Delta +V$ has  at least one discrete eigenvalue. Let $(\psi_{0\star},E_{0\star})$ denote the ground state eigenpair given by Theorem \ref{linear-gs}; here, $\psi_{0\star}$ may be chosen to be positive and normalized in $L^2$.
  Consider the equation for nonlinear bound states of NLS / GP with $g=-1$ (attractive nonlinear potential):
\begin{equation}
\left(\ -\Delta + V(x) - |\psi_E|^{2\sigma}\ \right)\psi_E\ =\ E\psi_E,\ \ E<0,\ \ \sigma\ge1.
\label{nlsgp-bdst}
\end{equation}
Then, there is a constant $\delta_0>0$  and a non-empty interval $\mathcal{I}=[E_{0\star}-\delta_0,E_{0,\star})$ such that for any $E\in\mathcal{I}$, equation \eqref{nlsgp-bdst} has a positive solution, which for
$E$ tending to $ E_{0\star}$ has the expansion:
\begin{equation}
\psi_E(x)\ =\ \rho(E) \left(\ \psi_{0\star}(x)\ +\ \mathcal{O}\left(\ \rho(E)^{2\sigma}\right)\  \right),\ \
 \rho(E)\ =\ \left|E_{0\star}-E\right|^{1\over2}\ \left(\ \int\psi_{0\star}^{2\sigma+2}\right)^{-\frac{1}{2\sigma}}
\label{psiEexpand}\end{equation}
\end{theo}

\subsection{Orbital stability of nonlinear bound states}\label{OrbitalStability}

In this section we discuss the {\it orbital Lyapunov stability} of nonlinear grounds states. In particular, we prove  nonlinear stability in $H^1(\R^n)$ modulo the natural group of symmetries.  We shall in this section consider  the case where $g=-1$, the case of focusing nonlinearity.
Given a positive $H^1$ solution of \eqref{nlsgp-bdst} introduce
its {\it orbit}:
\begin{align}
\cO_{\rm gs}\ &=\ \left\{\Psi_{E}(x)e^{i\gamma}\ :\ \gamma\in [0,2\pi)\right\},\ \ V\ne0\nn\\
\cO_{\rm gs}\ &=\ \left\{\Psi_{E}(x+x_0)e^{i\gamma}\ :\ \gamma\in [0,2\pi),\ x_0\in\R^n\right\},\ \ V\equiv0\ .
\label{eq:gsorbit}
\end{align}

 We say that the  ground state is {\it orbitally stable'} in $H^1(\R^n)$ if the following holds. If $\Phi(x,t=0)$ is close to element of $\cO_{\rm gs}$ in $H^1$ then  for all $t\ne0$,
$\Phi(x,t)$ is $H^1$ close to some (typically $t$ dependent) element of $\cO_{\rm gs}$.  In order to make this precise, we introduce a metric which measures the distance from an arbitrary $H^1$ function to the ground state orbit:
\begin{align}
{\rm dist}\left(u,\cO_{\rm gs}\right)\ =\
 \inf_\gamma\ \| u - \Psi_0e^{i\gamma}\ \|_{H^1},\ \ V\ne0\nn\\
 {\rm dist}\left(u,\cO_{\rm gs}\right)\ =\
 \inf_{\gamma,x_0}\ \| u(\cdot+x_0) - \Psi_0e^{i\gamma}\ \|_{H^1},\ \ V\equiv0.
 \label{eq:dist}
 \end{align}
A more precise statement of $H^1-$ Lyapunov stability is as follows. For any $\epsilon>0$ there is a $\delta>0$ such that if
\begin{equation}
{\rm dist}\left(\Phi(\cdot,0),\cO_{\rm gs}\right)\ <\ \delta
\label{eq:deltachoose}\end{equation}
then for all $t\ne0$
\begin{equation}
{\rm dist}\left(\Phi(\cdot,t),\cO_{\rm gs}\right)\ <\ \epsilon.
\nn\end{equation}

Nonlinear bound states are critical points of the Hamiltonian
 $\cH$ subject to fixed $L^2$ norm, $\cN$. In particular, a nonlinear bound state, $\psi_E$, of frequency $E$ satisfies $ \delta\cE_E[f,f^*]/\delta f^*\ =\ 0$,
 where
  \begin{align}
  \cE_E[f] &=  \cH[f]\ -\ E\int |f|^2\nn\\
 \cH[f] &\equiv \int \left|\nabla f\right|^2+V(x)|f|^2 - \frac{1}{\sigma+1}|f|^{2\sigma+2}
 \nn\end{align}
 For subcritical nonlinearities, $\sigma<2/d$, stable nonlinear ground states may be realized as constrained global minimizers of the variational problem:
 \begin{equation}
  \inf_{f\in H^1} \cH[f]\qquad \textrm{ subject to fixed}\ \ \cN[f]\ >\ 0.
  \label{nl-var}
  \end{equation}
   Orbital $H^1$ stability is a consequence of the compactness properties of arbitrary minimizing sequences \cite{CL:82,Lions:84}.
   \medskip

   However in general, depending on the details of the nonlinear terms, stable solitary waves may arise as  {\it local} minimizers of $\cH$ subject to fixed $\cN$.  We now discuss stability, in this more general setting. Introduce the linearized operators $L_\pm$, which are real and imaginary parts of the second variation of the energy $ \mathcal{E}_E$:
   \begin{align}
 L_+ \ &\equiv\ -\Delta\ +\ V(x)\ -\ (2\sigma+1) \psi_E^{2\sigma}\ -\ E .
 \label{L+def}\\
 L_- \ &\equiv\ -\Delta\ +\ V(x)\ -\ \psi_E^{2\sigma}\ -\ E
 \label{L-def}
 \end{align}

 We define the {\it index of $\psi_E$} by
 \begin{equation}
  {\rm index}\left(\psi_{E}\right)\ = \textrm{number  of strictly negative eigenvalues of}\   L_+\ .
  \label{index}\end{equation}

\begin{theo}[$H^1$ Orbital Stability]\label{w-gss}
\begin{enumerate}
\item Assume the conditions\\

(S1)\ Spectral condition:\
 $\psi_E>0$ and ${\rm index}\left(\psi_E\right)=1$

(S2) Slope condition
\begin{equation}
\frac{d}{d E} \cN[\psi_E]\ \equiv\ \frac{d}{d E} \int\ |\psi_E(x)|^2 dx\ <\ 0
\label{slope}\end{equation}
Then,  $\psi_E$ is $H^1$ orbitally stable.
 \item Assume $\psi_E>0$.  If  ${\rm index}\left(\psi_E\right)\ge2$ or $\frac{d}{d E} \cN[\psi_E]>0$ then $\psi_E$ is linearly exponentially unstable. That is, the linearized evolution equation (see \eqref{lin-ham} below) has a spatially localized solution which grows exponentially with $t$.
\end{enumerate}
\end{theo}
See \cite{Weinstein:85,Weinstein:86,GSS:87,Jones:88,Grillakis:90,SFIW:08}. In \cite{VK:73} it was shown that $\D_E\mathcal{N}[\psi_E]>0$ implies the existence of an exponentially growing mode of the linearized evolution equation.\medskip

We give the idea of the proof of stability. For simplicity, suppose $V(x)$ is non-trivial, In this case
the ground state orbit consists of all phase translates of $\psi_E$; see the first line of
 \eqref{eq:gsorbit}. Let $\epsilon$ be an arbitrary
 positive number. We have for $t\ne0$, by choosing $\delta$ in
  (\ref{eq:deltachoose}) sufficiently small
 \ba
 \epsilon^2\ &\sim&\ \mathcal{E}_{E}[\Psi(\cdot,0)]\ -\ \mathcal{E}_{E}[\psi_E]\nn\\
                &=&\ \mathcal{E}_{E}[\Psi(\cdot,t)]\ -\ \mathcal{E}_{E}[\psi_E],\
                 {\rm\ \ \  by\ conservation\ laws},\nn\\
                 &=&\  \mathcal{E}_{E}[\Psi(\cdot,t)e^{i\gamma}]\ -\ \mathcal{E}_{E}[\psi_E],\
                  {\rm\  \ \ \ by\ phase\ invariance}\nn\\
                  &=&\  \mathcal{E}_{E}[\psi_E+u(\cdot,t)+iv(\cdot,t)]\ -\
                  \mathcal{E}_{E}[\psi_E]\
                   {\rm (definition\ of\ the\ perturbation }\ u+iv,\ u,v\in\R)\nn\\
                   &\sim&\ \left\langle L_+u(t), u(t)\right\rangle\ + \ \left\langle L_-v(t), v(t)\right\rangle
                    \ ({\rm by\ Taylor\ expansion\ and\ } \delta
                    \mathcal{E}_{E}[\psi_E]=0)
                \label{eq:chain}
\ea
If $L_+$ and $L_-$
  were positive definite operators, implying the existence of positive constants $C_+$ and $C_-$ such that
  \begin{align}
  \left\langle L_+u, u\right\rangle\ &\ge\ C_+\ \|u\|_{H^1}^2,\label{eq:pd+}\\
    \left\langle L_-v, v\right\rangle\ &\ge\ C_-\ \|v\|_{H^1}^2\label{eq:pd-}
  \end{align}
  for all $u,v\in H^1$, then it would follow from (\ref{eq:chain}) that
  the perturbation about the ground state,\ $u(x,t)+iv(x,t)$, would
  remain of order $\epsilon$ in $H^1$ for all time $t\ne0$.
   The situation is however considerably more complicated. The relevant facts to note are as follows.
 \begin{itemize}
 \item[(1)] $L_-\psi_E=0$, with $\psi_E>0$. Hence, $\psi_E$ is the ground state of $L_-$,\  $0\in\sigma(L_-)$, and $L_-$ is a non-negative
 with continuous spectrum\ $[\ |E|,\infty)$.
 \item[(2)] For small $L^2$  nonlinear ground states,
  $L_+$ has exactly one strictly negative eigenvalue and continuous spectrum\ $[\ |E|,\infty)$.
  \end{itemize}
  The zero eigenvalue of $L_-$ and the negative eigenvalue of $L_+$ constitute two {\it bad} directions, which are treated as follows, noting
  that $u(\cdot,t)$ and $v(\cdot,t)$ are not arbitrary $H^1$ functions but
  are rather constrained by the dynamics of NLS.

  To control  $L_-$, we choose $\gamma(t)$ so as to minimize the distance of the solution to the ground state orbit, (\ref{eq:dist}). This yields the  codimension one contraint on $v$:\ $\left\langle v(\cdot,t),\psi_E\right\rangle=0$, subject to which (\ref{eq:pd-}) holds with $C_->0$.

  To control $L_+$, we observe that since the $L^2$ norm is invariant for solutions, we have the codimension constraint on $u$:\
   $\left\langle u(\cdot,t),\psi_E\right\rangle=0$. Although $\psi_E$ is not the ground state
   of $L_+$, it can be shown by constrained variational analysis that
   if the slope condition \eqref{slope} holds, then constraint on $u$ places $u$ in the positive cone of $L_+$, {\it i.e.}  $C_+>0$ in
   (\ref{eq:pd+}).
   Thus, positivity (coercivity) estimates (\ref{eq:pd+}) and (\ref{eq:pd-}) hold and $\mathcal{E}_{E}$ serves as a Lyapunov functional which controls the distance of the solution to the ground state orbit. The detailed argument is presented in \cite{Weinstein:86}.
   \medskip

   We also note that the role played by the second variation of $\mathcal{E}$ in the linearized time-dynamics \cite{Weinstein:85}. Let $\Psi(t)=(\psi_E+u+iv)e^{-iEt}$. The linearized perturbation,   $(u(t),v(t))^t$,  satisfies the linear Hamiltonian system:
   \begin{equation}
   \D_t \begin{pmatrix} u\\ v\end{pmatrix}\ =\ \begin{pmatrix} 0& L_-\\ -L_+& 0\end{pmatrix} \begin{pmatrix} u\\ v\end{pmatrix}\ ,
   \label{lin-ham}\end{equation}
   with conserved (time-invariant) energy
   \begin{align}
   \mathcal{Q}(u,v)&\equiv
   \left\langle L_+u, u\right\rangle\ + \ \left\langle L_-v, v\right\rangle
   \label{lin-en-def}\end{align}
The above constrained variational analysis underlying nonlinear orbital theory
 corresponds to the $H^1$ boundedness of the linearized flow \eqref{lin-ham}, restricted
 on a finite codimension subspace. This subspace is expressible in terms of symplectic orthogonality
 to first order generators of symmetries acting on the ground state. This structure is used centrally in many works on nonlinear asymptotic stability and nonlinear scattering theory of solitary waves; see section \ref{gss-ep} and references cited.

 \medskip

In the following subsections, we give several examples: a) the free NLS soliton, b) the nonlinear defect mode of a simple potential well, c) nonlinear bound states and symmetry breaking for the double well, and
d) {\it gap solitons} of NLS / GP with a periodic potential.

\subsection{The free soliton of focusing NLS: $V\equiv0$ and $g=-1$}\label{free-soliton}

 In this case, there is a unique (up to spatial translation)
positive and symmetric solution, a ground state, which is smooth and decays exponentially as $|x|\to\infty$.
For results on existence, symmetry and uniqueness of the ground state see, for example, \cite{Strauss:79,Berestycki-Lions:83,Gidas-Ni-Nirenberg:79,Kwong:89}.

 For example, the focusing one-dimensional cubic ($\sigma=1$) nonlinear Schr\"odinger equation
 \begin{equation}
 i\D_t\psi\ =\ - \D_x^2\psi\ -\ \abs{\psi}^2\psi,\label{1d-nls}\end{equation}
 has the solitary standing wave:
\begin{align}
 \psi_{sol}(x,t)\ &=
e^{i t}\sqrt{2}\ \sech\left(x\right)\ ,\label{1d-soliton} \end{align}

By the symmetries of NLS (section \ref{nls-nlsgp}), we have the extended family
 of solitary traveling waves of arbitrary negative frequency $E=-\lambda^2,\ \lambda>0$:
 \begin{align}
 & {\cal G}_{\lambda,x_0,v,\theta}\left[\psi_{sol}\right](x,t)\ =\
   \lambda e^{i\lambda^2 t} \sqrt{2}\ \sech\left(\lambda\ (x-x_0-2vt)\right)\ e^{iv(x-x_0-vt)}\ e^{i\theta};\ \
      v,\theta,x_0\in\R.
 \label{1d-soliton-fam}\end{align}

 Theorem \ref{w-gss} implies that the positive solitary standing wave of translation invariant NLS is orbitally stable if $\sigma<2/d$
  and is unstable if $\sigma\ge2/d$ \cite{Weinstein:83,Weinstein:86,GSS:87}. The solid curve in Figure \ref{L2vFreq} shows the family of solitary waves, \eqref{1d-soliton-fam}, of \eqref{1d-nls} bifurcating from the zero solution at the edge-energy of the continuous spectrum of $-\D_x^2$.

\subsection{$V(x)$, a simple potential well; model of a pinned nonlinear defect mode}\label{simple-defect}


In \cite{RW:88} the bifurcation of nonlinear bound states of NLS/GP from localized eigenstates of the linear Schr\"odinger operator $-\Delta+V$ was studied. The simplest case of such a {\it nonlinear defect mode} is for NLS / GP with $V(x)$ is taken to be a Dirac delta function potential well:
\begin{align}
i\D_t\psi\ &=\ H\psi - \abs{\psi}^2\psi,\ \ H=-\D_x^2\ -\ \gamma\d(x),\ \ \gamma>0\ .\label{nls-defect}
\end{align}
For the well-posedness theory of the intial-value-problem for \eqref{nls-defect}
 in $C^0([0,T);H^1(\R))$
see \cite{Jackson-Weinstein:04} and, more specifically, section 8D of \cite{DMW:11}.
In this case, the nonlinear defect mode of frequency $E=-\lambda^2<-(\gamma/2)^2$ is explicitly given by:
\begin{align}
\psi_{sol}(x,t;\gamma)\ &=
\lambda e^{i\lambda^2 t}  \sqrt{2}\ \sech\left(\lambda\abs{x} +
                            \tanh^{-1}\frac{\gamma}{\lambda}\right)\ e^{i\theta}
\nn \end{align}

\begin{figure}[ht!]
\label{L2vFreq}
\begin{center}
\includegraphics[width=3in]{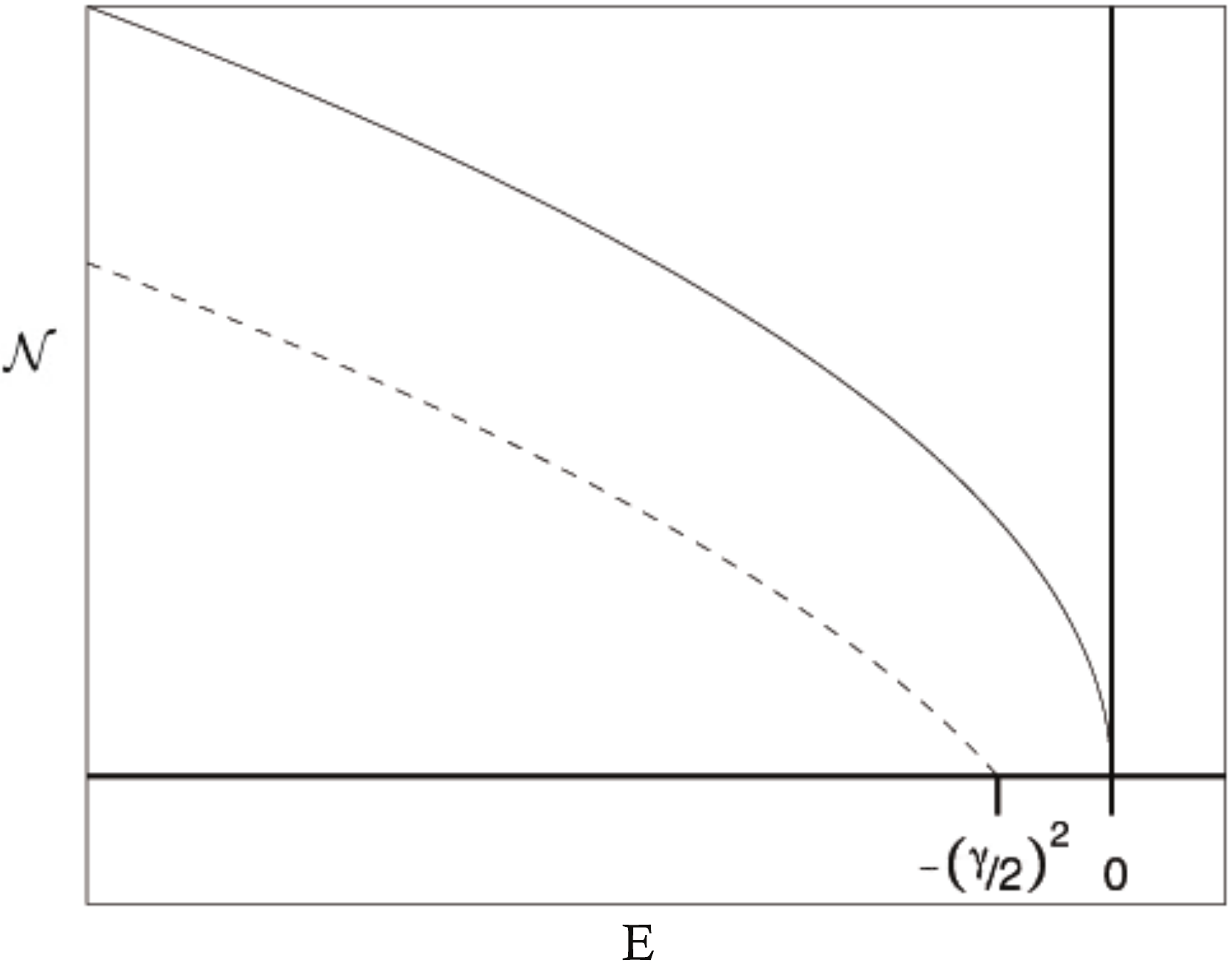}
\end{center}
\caption{{\footnotesize{Squared $L^2$ norm, $\mathcal{N}$,  vs. frequency for the free soliton (solid) and pinned soliton (dashed).}}}
\end{figure}

\nit This family of nonlinear bound states, which is pinned to the ``defect'' at $x=0$,  bifurcates from the zero solution at $E=E_\star=-(\gamma/2)^2$, the unique negative eigenvalue of $H$; see the dashed curve in Figure \ref{L2vFreq}.

\subsection{NLS / GP: Double well potential with separation, $L$  }\label{double-well}

\begin{figure}[ht]
\begin{center}
\begin{tabular}{cc}
\includegraphics[height=8cm]{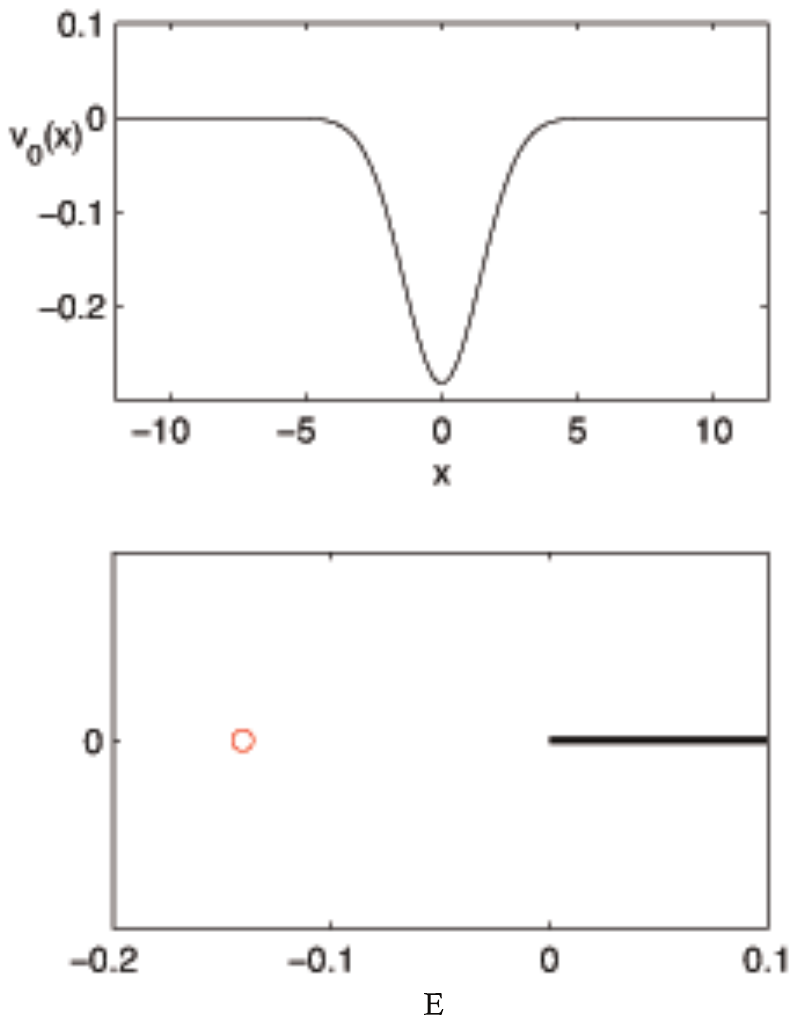}
\includegraphics[height=8cm]{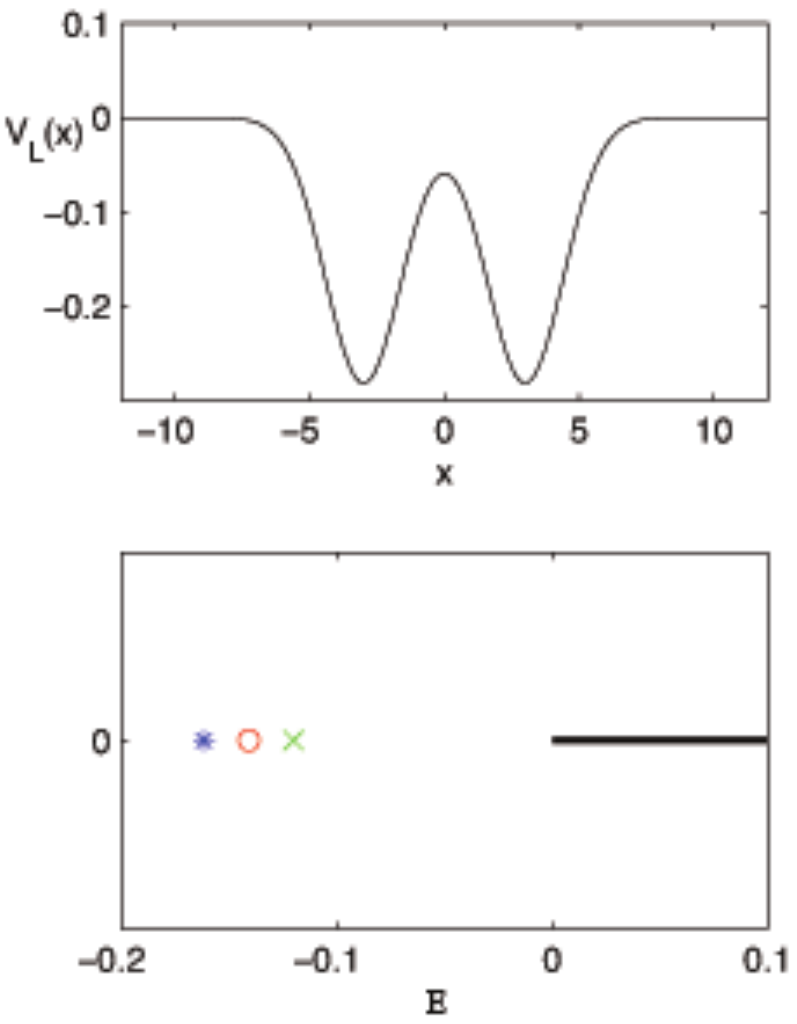} \\
\end{tabular}
\end{center}
\caption{{\footnotesize{Left: Single-well potential and its spectrum: one discrete eigenvalue, marked by ``${\rm  o}$'', and continuous spectrum, $\R_+$. Right: Double-well potential and its spectrum: two nearby discrete eigenvalues (marked ``$*$'' and ``$\times$'') and continuous spectrum,
$\R_+$ }}}
\label{s&d-wells}
\end{figure}

Consider now the NLS / GP with a double-well  potential, $V_L(x)$:
\begin{equation}
i\D_t\psi\ =\ - \Delta\psi\ -\ V_L(x) \psi - \abs{\psi}^2\psi \ .
\nn\end{equation}
We take the double-well, $V_L(x)$ to be constructed by centering  two identical single-bound state potentials, $V_0(x)$, of the sort considered in the example of section \ref{simple-defect},  at the positions $x=\pm L$:
\begin{equation}
V_L(x)\ =\  V_0(x+L)+V_0(x-L)
\label{VL-def}
\end{equation}
A sketch of a one-dimensional double well potential, $V_L(x)$, and the associated spectrum of $H=-\D_x^2+V_L(x)$ is displayed
in the right panels of Figure \ref{s&d-wells}.

\begin{figure}[ht]
\begin{center}
\includegraphics[height=5cm]{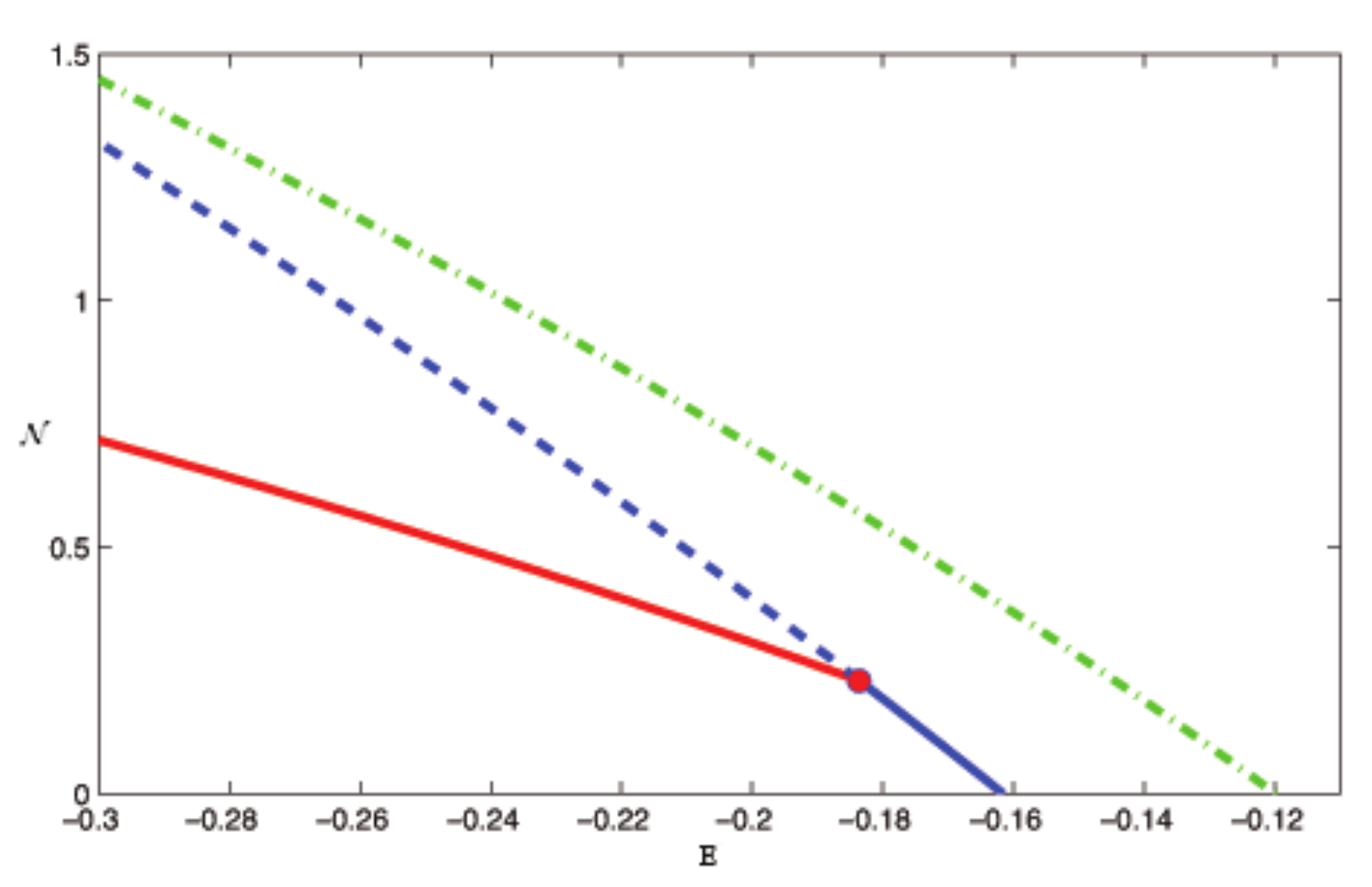}
\end{center}
\caption{{\footnotesize{Bifurcations from the two
 discrete eigenvalues of the double-well:
Bifurcation curve emanating from the zero state at the ground state (lowest) energy shows symmetry breaking at some positive $\cN_{cr}$.
Bifurcation curve emanating from the zero state at the excited state energy shows no symmetry breaking. }}}
\label{bif-curves}
\end{figure}

As in the example of section \ref{simple-defect} (see \cite{RW:88}) there are branches of nonlinear bound states which bifurcate from the zero solution at the discrete eigenvalue energies.
We focus on the branch emanating from the zero solution at the ground state eigenvalue of $H_L$. For small
 squared $L^2$ norm, $\cN$, this is the unique (up to the symmetry $\psi\mapsto e^{i\theta}\psi$) nontrivial solution branch.
 This solution has the same symmetries as the ground state of the linear double well potential \cite{Harrell:80}. That is, for $\cN$ small,  $\psi_{E(\cN)}$ is bi-modal with peaks centered at $x\pm L$. Increasing the  $L^2$ norm (or its square, $\cN$) we find that there is a critical value  $\cN_{\rm cr}>0$, such that
 for $\cN>\cN_{\rm cr}$ there are multiple nonlinear bound states; see Figure  \ref{bif-curves}. In particular, at $\left(E(\cN_{\rm cr}),\psi_{\cN_{\rm cr}}\right)$ there is a symmetry breaking bifurcation. Specifically, for $\cN>\cN_{\rm cr}$, there are three families of solutions:  the continuation of the symmetric branch (dashed curve continuation of the symmetric branch) and two branches of asymmetric states, corresponding to solutions whose mass is concentrated on the left or right side wells. The bifurcation diagram in Figure \ref{bif-curves} show only two branches beyond
 the bifurcation point. The solid leftward branching curve represents both asymmetric branches, one set of states being obtained from the other via a reflection about $x=0$.

 Figure \ref{bif-curves} also encodes  stability properties. For $0<\cN<\cN_{\rm cr}$, the (symmetric) ground state is stable, while for $\cN>\cN_{\rm cr}$ the symmetric state is unstable. at $\cN=\cN_{\rm cr}$ stability is transferred to the asymmetric branches (solid curve for $\cN>\cN_{\rm cr}$).

The above discussion is summarized in the following  \cite{KKP:10,KKSW:08}:

\begin{theo}[Symmetry Breaking Bifurcation]\label{symm-brk}
Consider the nonlinear eigenvalue problem
\begin{equation}
(-\Delta +V_L)\psi_E + g|\psi_E|^2\psi_E\ =\ E\psi_E,\ \ \psi_E\in H^1(\R^d)
\label{psiE-VL}
\end{equation}
where  $g<0$ and  $V_L$ denotes a double well potential with separation paramater, $L$, as in \eqref{VL-def}. Denote the ground and excited state eigenvalues of $-\Delta+V_L$ by:
\begin{equation}
E_{0\star}(L)\ <\ E_{1\star}(L)\ <\ 0.
\label{E01star}
\end{equation}
Let $\cN[f]=\int_{\R^d}|f|^2$.

There exists a positive constant,  $L_0$, such that
the following holds. For all $L\ge L_0$, there exists $\cN_{\rm cr}=\cN_{\rm cr}(L)>0$ such that
\begin{itemize}
\item[(a)]\ For any $\cN<\cN_{cr}$ there is a  unique non-trivial symmetric state.
\item[(b)]\ The point $(E,\psi_{E})=(E_{\rm cr},\psi_{E_{cr}})$  is a bifurcation point {\it i.e.}
\\ there are, for $\cN>\cN_{cr}$,  two bifurcating branches \underline{asymmetric} states, consisting  respectively of states concentrated about $x=\pm L$.
\item[(c)] \ Concerning the stability of these branches:
\begin{itemize}
\item[(c1)] $\cN<\cN_{cr}$:\ \ The symmetric branch is orbitally stable.
\item[(c2)] $\cN>\cN_{cr}$: The symmetric branch is linearly exponentially unstable;\\ these states have index $=2$; see \eqref{index}.
\item[(c3)] $\cN>\cN_{cr}$:\ The asymmetric branch is orbitally stable.
\end{itemize}
\item[(d)]  Symmetry breaking threshold:
\begin{equation} \cN_{cr}(L)\ \sim\ E_{1\star}(L)-E_{0\star}(L)\   \label{Ncr-estimate}
\end{equation}
(The gap, $E_{1\star}(L)-E_{0\star}(L)$ is exponentially small for large $L$ \cite{Harrell:80,Jackson-Weinstein:04}.)
\end{itemize}
\end{theo}

\begin{figure}[ht]
\begin{center}
\includegraphics[height=5cm]{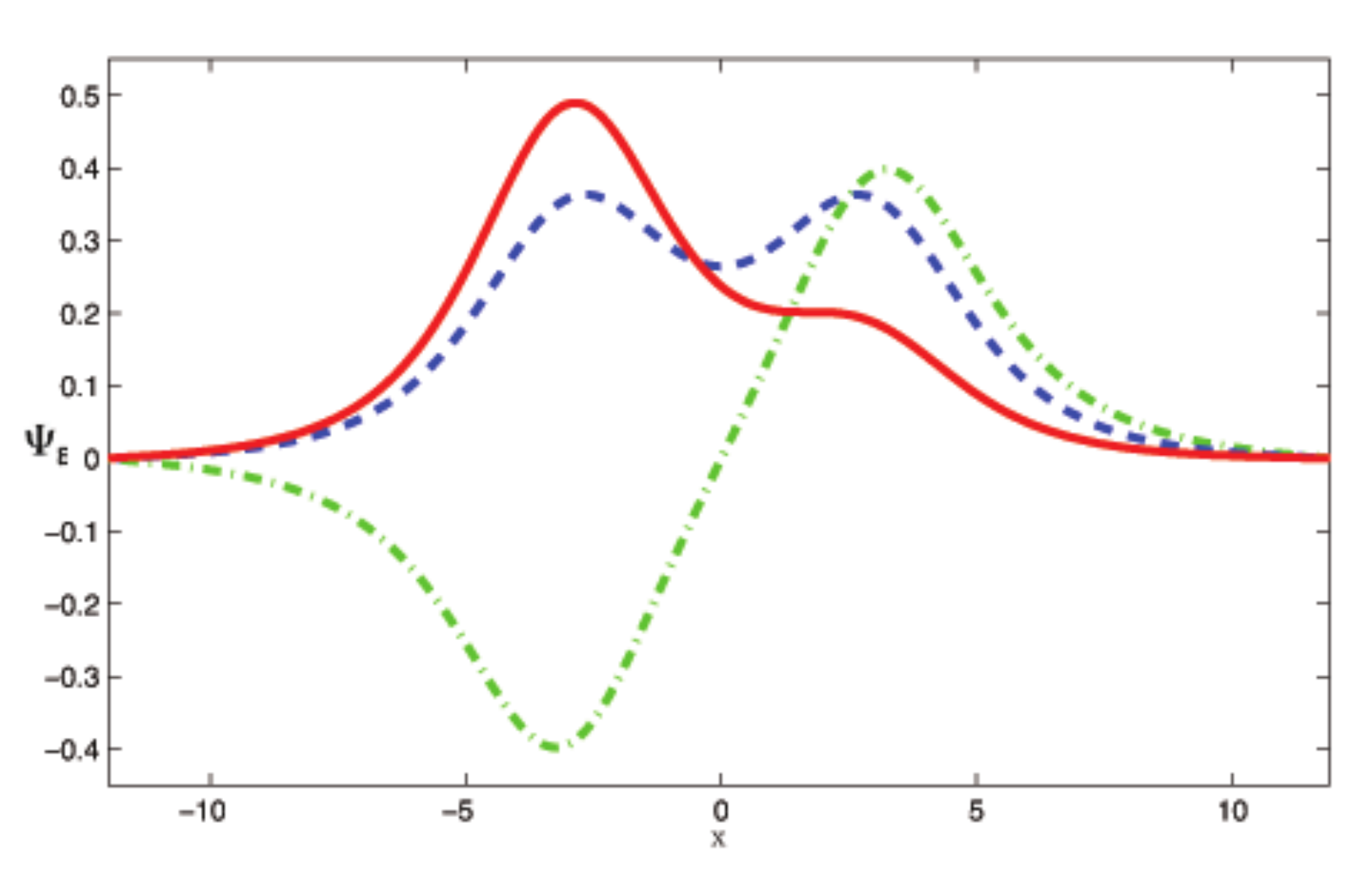} \\
\end{center}
\caption{{\footnotesize{Symmetric, asymmetric and anti-symmetric modes arising in the bifurcation diagram of  Figure \ref{bif-curves}.}}}
 \label{bif-modes}
\end{figure}

Theorem \ref{symm-brk} is proved by methods of bifurcation theory \cite{Golubitsky-Stewart-Schaefer:88,Nirenberg:74}. Specifically, for $L$ sufficiently large we can use a Lyapunov-Schmidt reduction
  strategy to reduce the nonlinear eigenvalue problem to a weakly perturbed system of two nonlinear algebraic equations depending on $\cN$.
The unknowns of this system are essentially the projections of the sought nonlinear bound state onto the  linear ground and excited states of $H_L$. The bifurcation structure of this pair of nonlinear algebraic equations is of the type displayed in Figure \ref{bif-curves}. This bifurcation structure is then shown to persist under the (infinite dimensional) perturbing terms to this finite algebraic reduction. The latter are controlled via PDE estimates and an implicit function theorem argument.

An illustrative study of the {\it global bifurcation structure} for the exactly solvable case where $V_L(x)$ is taken to be a sum of two attractive Dirac-delta wells centered at $x=\pm L$ is presented in
\cite{Jackson-Weinstein:04}.  In \cite{KK:06} the local bifurcation methods described above are used to study the NLS / GP with a triple well potential.
A recent study of  global properties of the bifurcating branches and their stability properties for general nonlinearities, $|\psi|^{2\sigma}\psi$, with $\sigma\ge1/2$  is presented in \cite{KKP:10}. In this work the authors prove, for the case of symmetric double-well potentials, that as $L$ varies a symmetry breaking bifurcation occurs once  the potential develops a local maximum at the origin. The analysis in \cite{KKP:10} allows for the bifurcation to occur at large ${\cal N}_{cr}$, outside the weakly nonlinear regime where the analysis in \cite{KKSW:08} applies.

Local bifurcation methods can be applied as well to study excited state branches. An example is the branch emanating from the linear excited state energy, along which it can be shown, for small $\mathcal{N}$,  that there are no secondary bifurcations; see  Figure \ref{bif-curves}  and the corresponding mode curve which changes sign in Figure
 \ref{bif-modes}.

An approach which complements the local bifurcation approach is variational. For example, in
\cite{AFGST:02}
symmetry breaking in a three-dimensional nonlinear Hartree model having an attractive non-local potential is proved for $\cN$ sufficiently large. Their arguments are straightforwardly adaptable to the variational formulation \eqref{nl-var} and yield that for $\sigma<2/d$ the ground state is asymmetric. The proof is based on a trial function argument and the intuition that for $\cN$ sufficiently large it is energetically preferable to concentrate mass in the left well or in the right well but not both.

\subsection{NLS / GP:  $V(x)$ periodic and the bifurcations from the spectral band edge}\label{spec-edge}

We now consider the nonlinear bounds states of NLS / GP for the case where $V$ is periodic on $\R^d$.
In this case, the spectrum of $H=-\Delta +V(x)$  is absolutely continuous.
Let $\psi_\mu$ denote a nonlinear bound state of frequency $\mu$.
 Any bifurcation of nontrivial solutions of \eqref{psiE-eqn} from the zero solution must occur at an energy in the continuous spectrum. Indeed, the example of the free soliton of section \ref{free-soliton} is an example; $V\equiv0$ is periodic (!) and has continuous spectrum equal to the half-line $\R_+=[0,\infty)$. The soliton bifurcates from the zero-amplitude solution (as measured say in $L^\infty$ or $L^2$) from the continuous spectral edge (solid curve in Figure \ref{bif-curves}).  Indeed, for NLS with $V\equiv0$ and  power nonlinearity $-|\psi|^{2\sigma}\psi$
 \begin{equation}
  \psi_{\mu}(x)=(-\mu)^{\frac{1}{2\sigma}}\psi_{-1}(\sqrt{-\mu} x)
  \label{soliton-scaling}
  \end{equation}
and therefore
   \begin{equation}
  \cN[\psi_{\mu}]\ =\  \|\psi_{\mu}\|_{L^2}^2\ =\ (-\mu)^{{1\over\sigma}-{d\over2}}\ \|\psi_{-1}\|_{L^2}\ .
  \label{N-scaling}\end{equation}

 For $V(x)$ a general periodic potential the spectrum of $H=-\Delta+V$ is the union of spectral bands obtained as follows. Consider the $d-$ parameter family of periodic elliptic eigenvalue problems:
 \begin{align}
& \left(\ -\left(\nabla+ik\right)^2\ +\ V(x)\ \right)u(x)\ =\ \mu\ u(x)\nn\\
 & u(x)\ \textrm{ is periodic with the periodicity of} \ V(x)\ ,\label{fb-evp}
 \end{align}
 where $k$ varies over a fundamental dual lattice cell, the Brillouin zone, $\mathcal{B}\subset\R^d$.
For each fixed $k\in\mathcal{B}$, the spectrum of \eqref{fb-evp} is discrete and consists of eigenvalues denoted:
 \begin{equation}
 \mu_1(k)\le\mu_2(k)\le\cdots\le\mu_b(k)\le\cdots\ ,
 \end{equation}
 listed with multiplicities and tending to positive infinity. The corresponding eigenfunctions are denoted, $p_j(x;k)$. The states $\{e^{ikx}p_j(x;k)\}$ where $j\ge1$ and $k\in\mathcal{B}$ are complete in $L^2(\R^d)$.

  It can be shown that nonlinear bound states bifurcate from edges of the spectral bands: to the left, if the nonlinearity is attractive ($g<0$) and to the right if the nonlinearity is repulsive ($g>0$) \cite{Ilan-Weinstein:10,Kivshar-Pelinovsky:04,Shi-Yang:07}. We expect similar scaling of the $L^2$ norm of such more general edge-bifurcations:
   \begin{equation}
  \cN[\psi_{\mu}]\ =\  \|\psi_{\mu}\|_{L^2}^2\ \sim \ |\mu-\mu_\star|^{{1\over\sigma}-{d\over2}}\
  \label{N-gap-scaling}\end{equation}
  where $|\mu-\mu_\star|$ is the distance to the spectral band edge located at $\mu_\star$.
  Now the details of the bifurcation at the edge depend on the periodic structure. We first give a heuristic picture and then state a precise theorem.

   For $|\mu-\mu_\star|$ small, that is for $\mu$ near a spectral band edge, the nonlinear bound state, $\psi_\mu(x)$, should be exponentially localized with decay $\psi_\mu(x)\sim \exp(-|\mu-\mu_\star|^{1\over2}|x|)$ for $|x|\to\infty$. Due to the separation of length scales:
   $|\mu-\mu_\star|^{-1}\gg\ \textrm{period of V}$, we expect $\psi_\mu(x)$ should oscillate like the Floquet-Bloch mode $p_1(x;0)=p_\star(x)$, associated with the band edge energy, $\mu_\star$, with a slowly varying and spatially localized amplitude, $F(x)$,  :
\begin{equation}
\psi \sim\ \delta^{{1\over\sigma}}F(\delta x)\times p_\star(x),\ \ \delta= |\mu-\mu_\star|^{1\over2}\ ,
\label{edge-heuristics}
\end{equation}
where $F(x)$ satisfies  an effective medium (homogenized / constant coefficient) NLS equation.

To simplify the discussion we will focus on bifurcation from the lowest energy (bottom) of the continuous spectrum, $\mu_\star=\mu_1(0)$,  for the case of an attractive nonlinearity ($g<0$).  Before stating a precise result, we introduce the inverse effective mass matrix associated with the bottom of the continuous spectrum defined by
\begin{align}
\label{inv-eff-mass}
 A_{\rm eff}\equiv \frac{1}{2} D^2\mu_1(k=0)\ &=\ \frac{1}{2}\ \left(  \frac{\D^2 \mu_1(k=0)}{\D{k_i}\D{k_j}}\right)_{1\le i,j\le d}\nn\\
 & =\  \delta_{ij} - \frac{4\ \left\langle \D_{x_j}p_\star\ ,\  (-\Delta+V-\mu_\star)^{-1}\ \D_{x_i}p_\star
          \right\rangle}{\left\langle p_\star\ ,\ p_\star \right\rangle}
    \end{align}
and $\Gamma_{\rm eff}$, the effective nonlinearity coefficient defined by
\begin{equation}
\label{eff-nonlin}
\Gamma_{\rm eff}\ =\ \frac{\int\ p_\star(x)^{2\sigma+2}\ dx}{\int\ p_\star(x)^2\ dx}\ >\ 0.
\end{equation}
The matrix $A_{\rm eff}$, is clearly symmetric and, for the lowest band edge, it is positive definite \cite{Kirsch-Simon:87}.

The following result, proved in  \cite{Ilan-Weinstein:10}, describes the bifurcation
from the bottom of continuous spectrum (left end point of the first spectral band). See also \cite{Kivshar-Pelinovsky:04,Sparber:06,Shi-Yang:07}. A proof of the case of a general band edge is discussed in \cite{Ilan-Weinstein:10}.

\begin{theo}[Edge bifurcations of nonlinear bound states for periodic potentials]\label{edge-bifurcation}
Let $V(x)$ denote a smooth and even periodic potential in dimension $d=1,2$ or $d=3$.
 Let $x_0$ denote any local minimum or maximum  of  $V(x)$. Denote by $\mu_\star=\inf\sigma(-\Delta+V)$.
Let  $F_{A,\Gamma}(y)$ the unique, centered at $y=0$,  positive $H^1(\R^d)$ solution (ground state)
 of the effective / homogenized nonlinear Schr\"odinger equation with inverse effective mass matrix, $A$, and effective nonlinearity, $\Gamma$.
 \begin{align}
&      -\sum_{i,j=1}^d\ \frac{\D}{\D{y_i}} A_{{\rm eff},ij}\frac{\D}{\D{y_j}} F(y)\ - \G_{\rm eff}\ F^{2\sigma+1}(y)\ =\ - F(y)~.
\label{effective-soliton}
\end{align}

Then, there exists $\delta_0>0$ such that for all $\mu\in (\mu_\star-\delta_0, \mu_\star)$ there is a family of nonlinear bound states $\mu\mapsto \psi_\mu(x)$
\begin{align}
 \psi_\mu(x)\ -\ \left(\mu_*-\mu\right)^{1\over2\sigma}\ F_{A_{\rm eff},\Gamma_{\rm eff}}\left(\ \sqrt{\mu_*-\mu}\ (x-x_0)\ \right)\  p_\star(x)\ \to\ 0,\ \ {\rm as}\ \mu\uparrow \mu_\star\ {\rm in}\ H^1(\R^d)\ .
\label{edge-bif}
\end{align}
\end{theo}

The proof of Theorem \ref{edge-bifurcation} \cite{Ilan-Weinstein:10}, like the study of bifurcations from discrete eigenvalues, proceeds via Lyapunov-Schmidt reduction and application of the implicit function theorem. However, while the analysis of bifurcation from discrete spectrum leads to a finite dimensional (nonlinear algebraic) bifurcation equation, bifurcation from the continuous spectrum leads to an infinite dimensional bifurcation equation, a nonlinear homogenized partial differential equation \eqref{effective-soliton}; see also \cite{DPS:09b,DU:09}. Examples of recent applications of these and related ideas to other systems appears in  \cite{DVW:15,DVW-oscillatory:14,Jenkinson-Weinstein:14,FW:12,FW-Biarretz:12,FW:14,FLW-pnas:14,FLW:14}.

Since \eqref{effective-soliton} is a constant coefficient equation, $F_{A,\Gamma}$ can be related, via scaling, to the nonlinear ground state of the translation invariant nonlinear Schr\"odinger equation. Thus, the shape of the bifurcation diagram $\mu\mapsto\cN[\psi_\mu]$ (see Figure \ref{bif-curves}) can be deduced for $\mu$ near $\mu_\star$; see \cite{Ilan-Weinstein:10}.
 We also note that the stability / instability properties of ground states of \eqref{effective-soliton}
 for the time-dependent effective nonlinear Schr\''odinger equation, are a consequence of Theorem  \ref{w-gss} in the translation invariant case.

\begin{figure}\centering
\includegraphics[width=6.5cm,height=6cm]{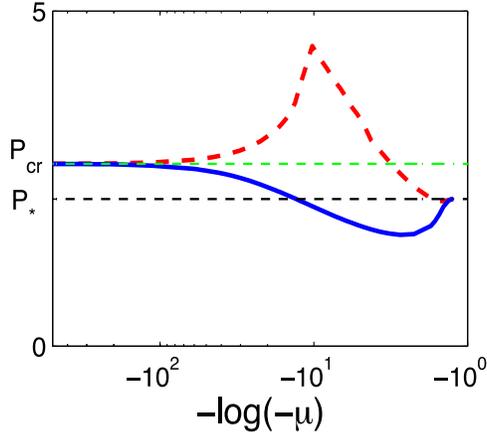}
\caption{{\footnotesize{$P=\cN[\psi_\mu]$ vs. $\mu$ for the 1d - NLS / GP with periodic potential, $V(x)$ for the case $\sigma=\sigma_{\rm cr}=2/d=2$. Dashed line $P=P_{\rm cr}$ is the ($\mu-$ independent) value of $\cN[\psi_\mu]$ for the case $V\equiv0$. Solid curve is the curve $\mu\mapsto\cN[\psi_\mu]$, where $\mu<\mu_\star=\inf(\sigma(-\Delta+V))$ for the case where $V$ is a non-trivial periodic potential. Nontrivial effective mass matrix implies $P_\star=\lim_{\mu\uparrow\mu_\star}\cN[\psi_\mu]<P_{\rm cr}$. See \cite{Ilan-Weinstein:10}. }}}
\end{figure}

\section{Soliton / Defect Interactions}\label{soliton-defect-intrxns}

In this section we turn to the detailed time dynamics of a soliton-like nonlinear bound state interacting with a potential. This question has fundamental and applications interest. From the fundamental perspective it is an important problem in the direction of developing a nonlinear scattering theory for non-integrable Hamiltonian PDEs.  And, from an applications perspective, nonlinear waves often arise in systems with impurities, {\it e.g.} random defects in fabrication or those deliberately inserted into the medium to influence the propagation;
 see, for example, \cite{GSW:02}.

Referring to Figure \ref{gscapture} we sketch the different stages present in the dynamics of a soliton which is initially incident
 upon a potential well:

\begin{figure}[h]
\vspace{-.125in}
\begin{center}
\includegraphics[width= 4in,height=2in,angle=0]{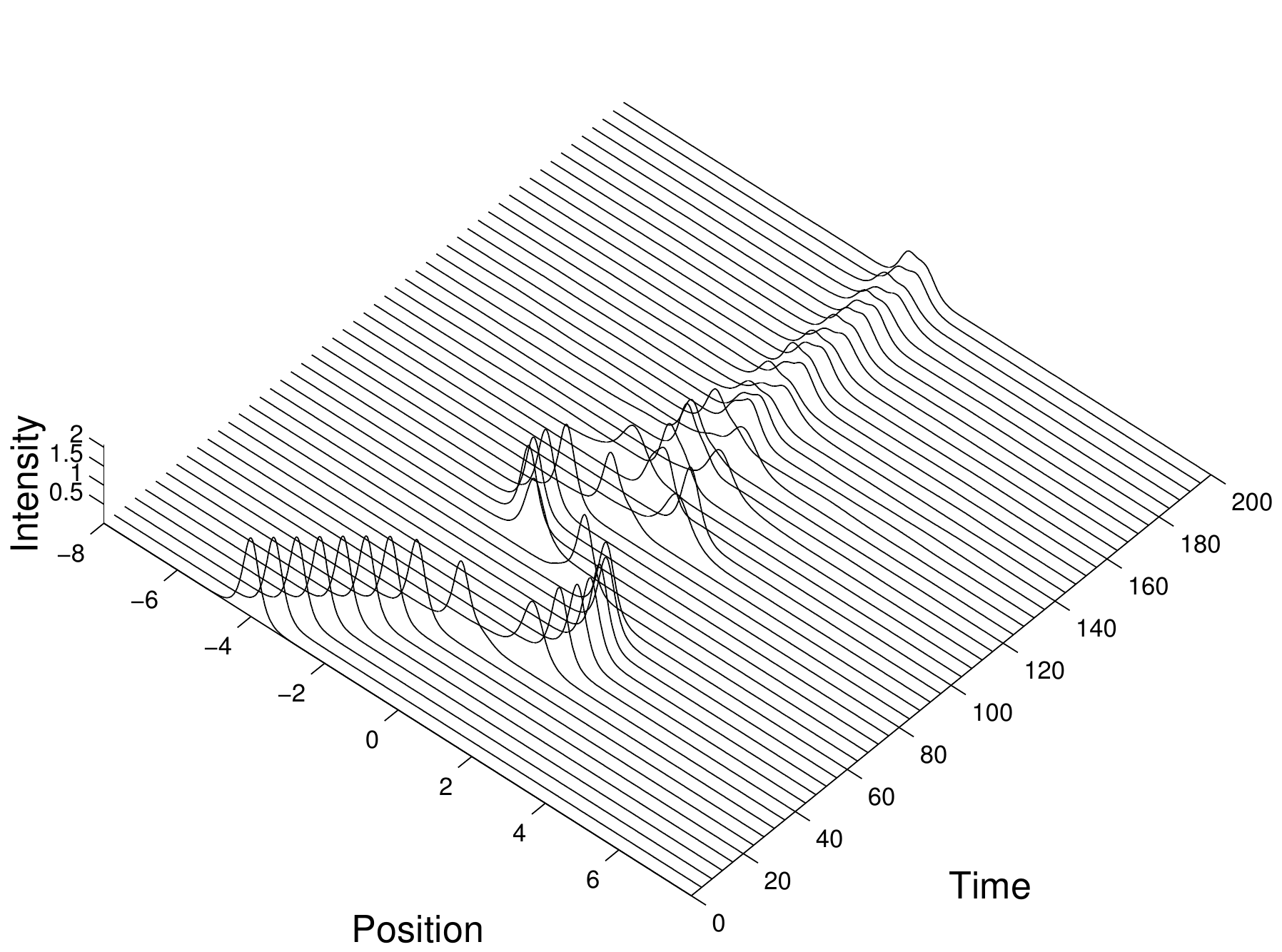}
\end{center}
\label{gscapture}
\caption{{\footnotesize{A soliton interacts with a defect, support around $x=0$. As time advances, the coherent structure interacts with the defect. Part of the incident energy is scattered to infinity and part of it is trapped in the defect. The trapped energy settles down to a stable nonlinear defect mode.}}}
\end{figure}

The results of formal asymptotic and numerical studies of soliton / defect interactions (see, for example, \cite{GSW:02,GHW:04,CM:95}) suggests a
description of the time-dynamics in terms of overlapping time epochs. Very roughly, these are:\\
(i)\  short and long / intermediate time-scale {\it classical particle-like dynamics}, which govern motion and deformations of the soliton as it interacts with the defect and exchanges energy with the defect's internal modes (pinned nonlinear defect modes) and \\
(ii)\ the very long time scale, during which the system's energy resolves into outgoing solitons moving away from the defect, small amplitude waves which disperse to infinity and an asymptotically stable nonlinear ground state of supported in the defect. A key process in this {\it asymptotic resolution} is radiation damping due to coupling of discrete and continuum degrees of freedom. See sections \ref{res-rad-damp} and \ref{gss-ep}.

Examples of rigorous analyses for {\it transient / intermediate time-scale}  regimes:
\begin{itemize}
\item[(i)]\  {\it Soliton scattering from a potential well}: Detailed reflection, transmission and trapping for solitons incident on a potential barrier or potential well  \cite{HMZ:07a,HMZ:07b,HZ:07,Datchev-Holmer:09}.
\item[(ii)]\ {\it Soliton evolving in a potential well}: In  \cite{HZ:08} the evolution of an order one soliton in a potential well is considered. The detailed {\it nonlinear breathing dynamics}, as the soliton relaxes toward its asymptotic state are considered in \cite{HZ:09};
see also section \ref{gss-ep} and the related work \cite{FGJS:04,Gang-Sigal:06,Gang-Sigal:07}.\medskip

In \cite{MW:10} the evolution of a weakly nonlinear (small amplitude) NLS / GP solutions in a double-well, for which there is symmetry breaking (see section \ref{double-well}) is studied.  Figure \ref{c-of-m-reduced} displays phase portrait of the reduced Hamiltonian dynamical system  in \cite{MW:10}. Orbits around the left (respectively, right) equilibrium map to a soliton executing a long-time nearly periodic (back and forth) motion with  the left (respectively, right) well of the double well. This analysis has been recently extended to certain orbits outside the separatrix \cite{GMW:13}.  \begin{figure}
\begin{center}
 \includegraphics[width=3in]{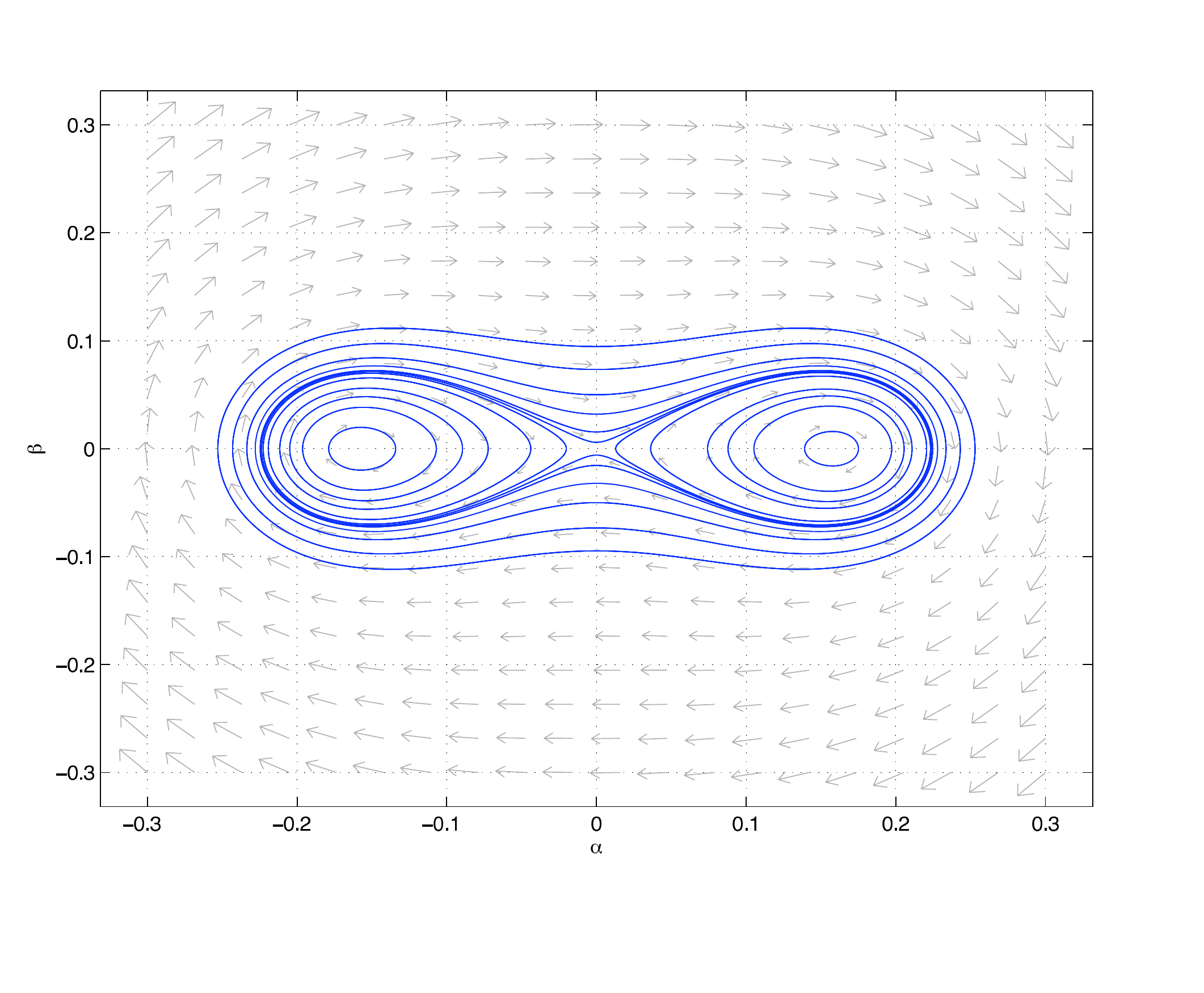}
\caption{{\footnotesize{Periodic dynamics of center of mass dynamics in the reduced (approximate) finite-dimenstional  Hamiltonian system \cite{MW:10}.  Equilibria corresponding to stable asymmetric states centered on left and right  local minima of double well. Oscillatory orbits decay to stable equilibria for the full, infinite-dimenstional NLS/GP. }}}
\end{center}
\label{c-of-m-reduced}
\end{figure}

For $t\gg1$ the center of mass eventually crystallizes on a stable nonlinear ground state. For $\cN>\cN_{\rm cr}$ this asymptotic state will be an asymmetric state (section \ref{double-well}) centered in the left or in the right well. See Figure \ref{c-of-m-full} which shows the computed center of mass motion for a solution of NLS / GP; initially there is oscillatory motion among the left and right wells. However, during each cycle some of the soliton's energy radiated away to infinity. The corresponding motion in the reduced phase portrait   Figure \ref{c-of-m-reduced} is damped and the actual center of mass trajectory is transverse to the level energy curves. Eventually the separatrix is crossed and the solution settles down to
a stable asymmetric state \cite{CM:11,SW:04,SW:05} We emphasize that this picture is only heuristic although there has been considerable progress toward understanding the radiation damping and ground state selection in such systems. We now turn to these phenomena in the next section. We conclude this section noting work on the dynamics of forced / damped NLS and its reduced finite dimensional dynamics, capturing regular and chaotic behavior; see, for example, \cite{CMM:02,SR-K:10}.

\begin{figure}
\begin{center}
\includegraphics[width=3in]{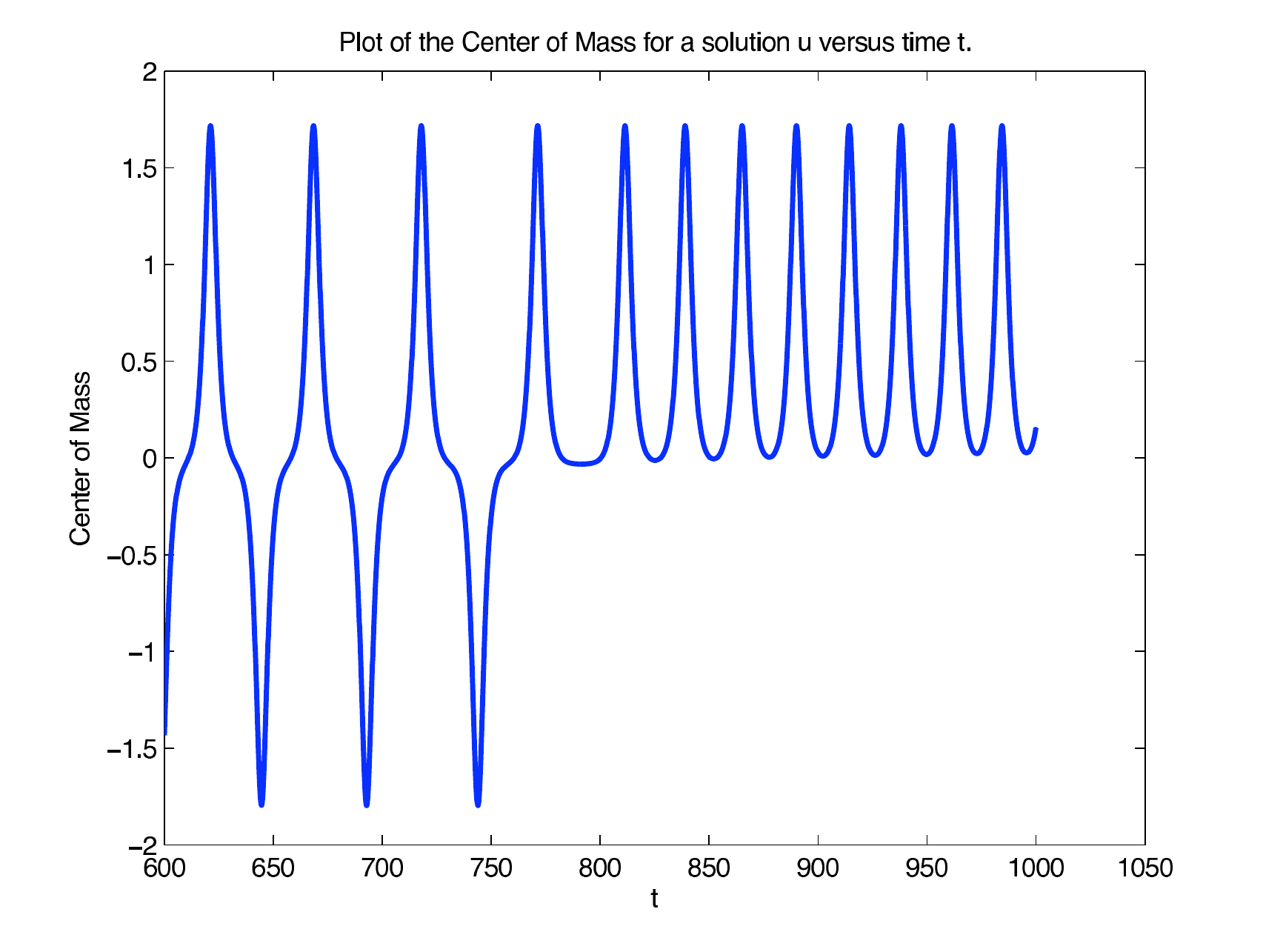} \\
\caption{{\footnotesize{Numerically computed dynamics of center of mass  in the NLS evolution \cite{MW:10}. Corresponding trajectory in the reduced phase portrait of Figure \ref{c-of-m-reduced} cycles around several times outside the separatrix, crosses the separatrix and then slowly spirals in toward the right stable equilibrium.}}}
\end{center}
\label{c-of-m-full}
\end{figure}

\end{itemize}

\section{Resonance, radiation damping and  infinite time dynamics}\label{res-rad-damp}

Our next goal is to discuss results concerning the infinite time dynamics of NLS / GP. We begin with an informal discussion of the linear Schr\"odinger equation (NLS / GP with $g=0$):

\begin{equation}
i\D_t\psi\ =\ ( -\Delta + V(x))\psi
\label{l-schrod}\end{equation}

Assume $V(x)$ is a smooth and sufficiently decaying real-valued  potentia for which
$-\Delta+V$ has bound states :
\begin{align}
&(-\Delta +V)\psi_j\ =\ \Omega_j \psi_j,\ \ j=0,1,\dots,\ \psi_j\in L^2\nn\\
& \psi(x,t)\ =\ e^{-i\Omega_j t}\ \psi_j(x)\nn
\nn\end{align}

Then, for sufficiently localized initial conditions, $\psi(0)$, the solution of the initial value problem for \eqref{l-schrod} may be written:
\begin{equation}
\psi(x,t)\ =\ e^{-i(-\Delta+V)t} \psi_0\ =\ \sum_j \left\langle \psi_j,\psi(0)\right\rangle e^{-i\Omega_jt}\psi_j(x) +\
R(x,t)\label{lin-sol}\end{equation}

The sum  in \eqref{lin-sol} is quasi-periodic while the second term, $R(x,t)$, radiates to zero in the sense that
 for some $\alpha>0$
\begin{equation}
\left\| R(\cdot,t)\right\|_{L^p(\R^d)}\ \lesssim\ t^{-\alpha},\ {\rm as}\ t\to\pm\infty
\end{equation}
Thus, as $t\to\infty$, the solution $\psi(x,t)$ tends to an asymptotic state which is quasi-periodic in time and localized in space.\medskip

\nit {\bf Question:} What is the large time behavior of solutions of the corresponding initial value problem for the nonlinear Schr\"odinger / Gross-Pitaevskii, $g\ne0$?
 \medskip

 For small amplitude it is natural to attempt expansion of solutions in the  basis of localized eigenstates and continuum modes, associated with  the unperturbed (solvable) linear Schr\"odinger equation.
In terms of these coordinates, the original  Hamiltonian PDE may be written as an equivalent dynamical system comprised of two weakly coupled subsystems:
\begin{enumerate}
\item[$\bullet$] a finite or infinite dimensional subsystem with {\bf discrete} degrees of freedom ( ``oscillators'' )
\item[$\bullet$] an infinite dimensional system (wave equation) governing a {\bf continuum} ``field''\ .
\end{enumerate}

These two systems are coupled due to weak nonlinearity. If one ``turns off'' the nonlinear coupling there are localized in space and time-periodic solutions, corresponding the eigenstates of the linear Schr\"odinger equation. If nonlinearity is present, new frequencies are generated, and these may lead to resonances among discrete modes or between discrete and continuum radiation modes. The latter type of resonance plays a key role in understanding the radiation damping mechanism central to  asymptotic relaxation of solutions to NLS / GP as $t\to\pm\infty$ and, in particular the phenomena of  {\it ground state selection} and {\it energy equipartition}.

These results will be discussed in section \ref{gss-ep}. A variant was studied in the context of the nonlinear Klein Gordon equation in \cite{SW:99,Bambusi-Cuccagna:11}. In the following two subsections we present a very simple example of emergent effective damping in an infinite dimensional Hamiltonian problem.

\subsection{Simple model - part 1:  Resonant energy exchange between an oscillator and a wave-field}\label{simple-1}

We consider a  solvable toy model of an infinite dimensional Hamiltonian system comprised of two subsystems, one governing discrete and the other governing continuum degrees of freedom. We shall see that coupling of these subsystems leads to energy transfer from the discrete to the continuum modes and the emergence of effective damping. Our model is a variation on the work of  Lamb (1900) \cite{Lamb:1900}  and   Weisskopf-Wigner (1930) \cite{Weisskopf-Wigner:1930}; see also \cite{SW-gafa:98,Costin-Soffer:01,SW-jsp:98,SW:99,Soffer-Weinstein:99,Kirr-Weinstein:01,Kirr-Weinstein:03,Kirr-Weinstein:05,CKP:06}

Consider a system which couples an  ``oscillator'' with amplitude $a(t)$ to a ``field'' with amplitude $u(x,t),\ t\in\R,\ \ x\in\R$:
\begin{align}
& \frac{ d a^\varepsilon(t)}{dt}\ +\ i\omega a^\varepsilon(t) \ =\ -\ \varepsilon u^\varepsilon(0,t),\label{oscillator}
&\\
&\D_tu^\varepsilon(x,t)\ +\ c\ \D_xu^\varepsilon(x,t)\ =\ \varepsilon\delta(x) a^\varepsilon(t).\label{field}
\end{align}
Here, $\varepsilon$ and $c$ are taken to be a real parameters, say with  $c>0$,
 and $\delta(x)$ denotes the Dirac delta function.
The amplitudes  $a(t)$ and $u(x,t)$ are complex-valued.

Note that the dynamical system \eqref{oscillator}-\eqref{field} conserves an energy, $\D_t\mathcal{E}[a(t),u(\cdot,t)]=0$.
\begin{align}
\mathcal{E}[a(t),u(\cdot,t)]\ &\equiv\  |a(t)|^2\ +\ \int_\R |u(x,t)|^2\ dx
\label{Edef}\end{align}
We consider the initial value problem for \eqref{oscillator}-\eqref{field} with initial data
\begin{equation}\label{data}
a(0)\ \textrm{arbitrary and}\ u(x,0)=0\ ;
\end{equation}
we perturb the oscillator initially, but not the field.

For $\varepsilon=0$ the  oscillator and wave-field are decoupled and there is an exact
time-periodic and finite energy global solution, a {\it bound state}:
\begin{equation*}
\left(\begin{array}{c} u^{\varepsilon=0}(x,t)\\ a^{\varepsilon=0}(t) \end{array}\right)\ =\
e^{-i\omega t}\ \left(\begin{array}{c} 0 \\  a(0) \end{array}\right).
\end{equation*}

For  $\varepsilon\ne0$, the oscillator and field are coupled; as time evolves  energy can be transfer
among the discrete and continuum degrees of freedom. It is simple to solve the initial value problem for any
choice of $\varepsilon$ and arbitrary initial data, {\it e.g.} by Laplace transform. In particular, one can proceed by solving the wave equation, \eqref{field}, for $u^\varepsilon=u(x,t;a^\varepsilon)$ as a functional of discrete degree of freedom $a^\varepsilon(t)$
 and then substitute the resulting expression for $u^\varepsilon\left(0,t;a^\varepsilon(t)\right)$ into the oscillator equation \eqref{oscillator}.
 The result is the closed oscillator
 \begin{align}
& \frac{ d a^\varepsilon(t)}{dt}\ +\ i\omega a^\varepsilon(t) \ =\ -\frac{\varepsilon^2}{2c} a^\varepsilon(t),\ \ t\ge0.
\label{closed-oscillator}
\end{align}
with an effective damping term, whose solution is:
\begin{equation}
a^\varepsilon(t)\ =\ e^{-i\omega t}\ e^{-\frac{\varepsilon^2}{2c}t}\ a_0\ =\ e^{-i(\omega-i\frac{\varepsilon^2}{2c})\ t}\ a_0,\ \ t\ge0.
\label{a-eps}
\end{equation}
 That is, in terms of the oscillator, the {\it closed} infinite dimensional conserved system for $a^\varepsilon(t)$ and $u^\varepsilon(x,t)$ may equivalently be viewed in terms of a reduced {\it open and damped} finite dimensional system for $a^\varepsilon(t)$.

Solving \eqref{closed-oscillator} for $a^\varepsilon(t)$ and substituting into the expression for $u^\varepsilon=u(x,t;a^\varepsilon)$, one obtains:
\begin{align}
& u^\varepsilon(x,t)\ = \left\{ \begin{array}{ll} \ 0,\ & x>ct\ge0  \nn\\
                          \frac{\varepsilon}{c}\ a_0\ e^{-i\omega(t-\frac{x}{c})}
\ e^{-\frac{\varepsilon^2}{2c}(t-\frac{x}{c})},\ &  x<ct
  \end{array}\right.
 \end{align}

To summarize, for $\varepsilon=0$, the decoupled system has a time-periodic finite energy bound state. For the coupled system, $\varepsilon\ne0$,  the bound state has a finite lifetime; it decays on a timescale of order  $\varepsilon^{-2}$. The bound state loses its energy to the continuum degrees of freedom; the lost energy is propagated to spatial infinity.
\bigskip

\subsection{ Simple model - part 2: Resonance, Effective damping and Perturbations of Eigenvalues in Continuous Spectra}\label{simple-2}

It's physically intuitive that an oscillator coupled to wave propagation in an infinite medium will damp.
 We now connect this damping to the classical notion of resonance.

We write the oscillator / wave-field model in the form
\begin{equation}
i\D_t\psi\ =\ A(\lambda)\ \psi,
\label{eq:model-schroed}
\end{equation}
where
\begin{equation}
\psi(t)\ =\
\left(
\begin{array}{c}
u(x,t) \\ a(t)
\end{array}
\right),\
A(\varepsilon)\ =\
\left(
 \begin{array}{cc}
 -ic\D_x  & +i\varepsilon\delta(x)\\
- i\varepsilon\left\langle \delta(\cdot) ,\cdot\right\rangle & \omega
 \end{array}
 \right)\
 \label{eq:psi-Aalpha-def}
\end{equation}

Let
\begin{equation}
\psi(t)\ =\ e^{-iEt}\psi_0\ =\ \left(
\begin{array}{c}
u(x,t) \\ a(t)
\end{array}
\right) =
e^{-iEt} \left(
\begin{array}{c}
u_0 \\ a_0
\end{array}
\right)
\end{equation}
This yields the spectral problem
\begin{equation}
A(\varepsilon)\psi_0=E\psi_0 \ ,
\label{eq:A_alpha-spec}
\end{equation}
or equivalently
\begin{equation}
E\left(
\begin{array}{c}
u_0 \\ a_0
\end{array}
\right)\ =\  \left(
 \begin{array}{cc}
 -ic\D_x& + i\varepsilon\delta(x)\\
 -i\varepsilon\left\langle \delta(y),\cdot\right\rangle & \omega
 \end{array}
 \right)\ \left(
\begin{array}{c}
u_0 \\ a_0
\end{array}
\right)\ .
\label{spectral-problem}
\end{equation}
%
This spectral problem may be considered on the Hilbert space:
\begin{equation}
\cH\ =\ \left\{\ \left(\begin{array}{c} u \\ a \end{array} \right)\in L^2(\R)\times\mathbb{C}\ :\ \|(u,a)\|_\cH<\infty  \right\}\ ,
\label{cHdef}\end{equation}
where
\begin{equation}
\|(u,a)\|_\cH= \int_\R |U(x)|^2 dx\ +\ |a|^2\ <\infty\
\label{cEdef}
\end{equation}

We note that
for $\varepsilon=0$ (decoupling of oscillator and wave-field), we have
 \begin{equation}
E\left( \begin{array}{c} u_0 \\ a_0 \end{array} \right)\ =\
\left( \begin{array}{cc} -ic\D_x  & 0\\  0 & \omega  \end{array}  \right)\
\left( \begin{array}{c}
u_0 \\ a_0
\end{array}
\right)
\end{equation}
$A(\varepsilon=0)$ is diagonal  and has spectrum given by:
\begin{itemize}
\item a point eigenvalue at $E=\omega$ with corresponding eigenstate
$\psi_\omega\ =\ \left(\begin{array}{c} 0 \\ 1\end{array}\right)$ and
\item continuous spectrum $\R=\{E=ck\ :\ k\in\R\}$ with eigenstates
$\psi_k\ =\ \left(\begin{array}{c} e^{ikx} \\ 0\end{array}\right),\
 k\in\R$\ .
 \end{itemize}
 Thus, $A(\varepsilon=0)$ has an embedded eigenvalue, $\omega$, in the continuous spectrum, $\R$, and
 the source the damping is seen to be this resonance. Understanding the coupling
 of oscillator and wave-field is therefore related to the perturbation
  theory  of an embedded (non-isolated) point in the spectrum; see, for example,
  \cite{RS4,Hislop-Sigal:96,SW-gafa:98} and references therein.

In this present simple setting, this perturbation problem can be treated as follows. Note that for $\varepsilon\ne0$, since $c>0$ we expect that energy is emitted by the oscillator and it gets carried to positive infinity through $u(x,t)$.
Thus we expect that on a fixed compact set, $u(x,t)$, will decay to zero. Alternatively, if we choose $\alpha>0$ and study $U(x,t)=e^{-\alpha x} u(x,t)$, $U(x,t)$  can be expected to decay to zero as $t$ advances. Indeed, it can be checked that $U(x,t)$ satisfies a {\it dissipative} PDE, for which the $L^2(\R)$ norm of $U(x,t)$ decays as $t$ increases; see the discussion of section \ref{simple-1}.

This behavior is reflected in the spectral problem. Consider the of change variables $(u,a)\mapsto (e^{\alpha x} U, a)$ in the spectral problem \eqref{spectral-problem}. This yields

\begin{equation}
E\left(
\begin{array}{c}
U_0 \\ a_0
\end{array}
\right)\ =\  \left(
 \begin{array}{cc}
 -ic\D_x - ic\alpha & + i\varepsilon e^{-\alpha x}\delta(x)\\
 -i\varepsilon\left\langle \delta(\cdot) e^{\alpha \cdot},\cdot\right\rangle & \omega
 \end{array}
 \right)\ \left(
\begin{array}{c}
U_0 \\ a_0
\end{array}
\right)\ =\ A_\alpha(\varepsilon)\  \left(
\begin{array}{c}
U_0 \\ a_0
\end{array}
\right),\
 \label{eq:H_alpha_spec}
\end{equation}
which when considered on $L^2(\R)$ is equivalent to the spectral problem \eqref{spectral-problem}  for $(u(x),a)$ in the weighted function space:

\begin{equation}
\cH_\alpha\ =\ \{(u(x),a)\ :\ \| u\|_{\cH_\alpha}<\infty\ \}, ,
\label{eq:cH_def}
\end{equation}
where
\begin{equation}
\ \ \| u\|_{\cH_\alpha}= \int_\R |u(x)|^2 e^{-2\alpha x} dx\ +\ |a|^2\ <\infty\ \ ,\qquad  \alpha>0.
\label{cEalpha}
\end{equation}

 \nit For $\varepsilon=0$,
 \begin{equation}
E\left(
\begin{array}{c}
v_0 \\ a_0
\end{array}
\right)\ =\
\left( \begin{array}{cc}
 -ic\D_x - ic\alpha & 0\\
           0& \omega
 \end{array}\right)
\left(
\begin{array}{c}
v_0 \\ a_0
\end{array}
\right)
\end{equation}
  $A_\alpha(0)$ is
  diagonal and has spectrum:
  \begin{itemize}
\item point eigenvalue at $E=\omega$ with corresponding eigenstate
$\psi_\omega\ =\ \left(\begin{array}{c} 0 \\ 1\end{array}\right)$
\item continuous spectrum in the lower half plane along the horizontal line $\{E=ck-ic\alpha\ :\ k\in\R\}$ with eigenstates
$\psi_k\ =\ \left(\begin{array}{c} e^{ikx} \\ 0\end{array}\right),\
 k\in\R$
 \end{itemize}
 Note that for $\alpha>0$, $\omega$ is an {\it isolated} eigenvalue of $A_\alpha(\varepsilon=0)$ and we can therefore implement a standard perturbation theory to calculate the effect of $\varepsilon\ne0$ on this eigenvalue.

 The first equation of (\ref{eq:H_alpha_spec}) is
 \begin{equation}
 \left(-ic\D_x-ic\alpha\right)v_0\ +\ i\varepsilon\delta(x)e^{-\alpha x}a_0\
  =\ Ev_0
  \label{eq:H_alpha1}
  \end{equation}
  This implies for $x\ne0$
  \begin{equation}
  v_0(x)\  =\ \left\{ \begin{array}{ll} e^{i(\frac{E}{c}x-\alpha)x}v_{0+} &\ x>0\nn\\
                                                     e^{i(\frac{E}{c}x-\alpha)x}v_{0-} &\ x<0
                            \end{array}
                            \right.
  \end{equation}
  Integration of (\ref{eq:H_alpha_spec}) over a small neighborhood of $x=0$ yields
  \begin{equation}
  v_{0+}-v_{0-}\ =\ +\frac{\varepsilon}{c}a_0
  \label{eq:jump0}
  \end{equation}
  The second equation of (\ref{eq:H_alpha_spec}) implies
  \begin{equation}
  (E-\omega)a_0\ =\ -\frac{i\varepsilon}{2}\left(\ v_{0+}+v_{0-}\ \right)
  \label{eq:avg0}
  \end{equation}
Now choose $v_{0+}\ne0$ and $v_{0-}=0$. Then,
 \begin{equation}
 E^\varepsilon\ =\ \omega\ -\ i\Gamma^\varepsilon,\ \ \Gamma^\varepsilon = \frac{\varepsilon^2}{2c}\ >\ 0\ .
 \label{eq:Eeqn}
 \end{equation}
 The corresponding eigenstate is
 \begin{equation}
 v^\varepsilon_{0,\alpha}(x)\ =\  \left\{\begin{array}{ll}
                      +\frac{\varepsilon}{c}a_0\ e^{i\omega\frac{x}{c}}
                       e^{ \frac{1}{c} (\frac{\varepsilon^2}{2c}-\alpha c)x},\ & x>0\\
                         0,\ & x<0\end{array}
                       \right.  \ \ ,
   \label{eq:v0+}
\end{equation}
which is in $L^2(\R)$ provided $\alpha >\frac{\varepsilon^2}{2c^2}$.
\medskip

For $\alpha=0$, we have
 \begin{equation}
 v^\varepsilon_{0}(x)\ =\  \left\{\begin{array}{ll}
                      +\frac{\varepsilon}{c}a_0\ e^{i\omega\frac{x}{c}}
                       e^{ \frac{1}{c} \frac{\varepsilon^2}{2c}x},\ & x>0\\
                         0,\ & x<0\end{array}
                       \right.  \ \ .
   \label{eq:v0+}
\end{equation}
The solution (\ref{eq:v0+}) satisfies an {\it outgoing radiation condition} at $x=+\infty$ and
 is in $L^2(\R;e^{-2\alpha x} dx)$ provided $\alpha >\frac{\varepsilon^2}{2c^2}$.
.

To summarize: the $\varepsilon=0$ eigenvalue problem has discrete eigenvalue, $E^0=\omega$, embedded in the continuous spectrum, $\R$ with corresponding eigenstate $\left(\begin{array}{c} 0 \\ 1\end{array}\right)$.  This eigenvalue perturbs, for $\varepsilon\ne0$, to a complex energy in the lower half plane, $E^\varepsilon=\omega-i\Gamma^\varepsilon,\ \Gamma^\varepsilon>0$ with corresponding eigenstate
 $\left(\begin{array}{c} v^\varepsilon_{0}(x) \\ 1\end{array}\right)$  which solves the eigenvalue equation with outgoing radiation condition at $x=+\infty$.
The complex energy, $E^\varepsilon$, may be viewed as a eigenvalue with normalizable eigenstate, $v_{0}(x)$ in the \underline{weighted} function space, $\mathcal{H}_\alpha$.
Since $E^\varepsilon$ is in the lower half plane, the corresponding time-dependent state is exponentially decaying as $t$ increases.

\begin{remark} Finally, we remark that the complex energy $E^\varepsilon$ may be viewed as a pole of the Green's function, when analytically continued from the upper half $E-$ plane to the lower half $E-$ plane.
\end{remark}

\section{Ground state selection and energy equipartition in NLS / GP}\label{gss-ep}

In this section we return to the question raised  in section \ref{res-rad-damp}, here restated.
 For simplicity we consider  NLS / GP  for the case with   cubic nonlinearity ($\sigma=1$):
\begin{align}
&i\D_t\Phi\ =\ \left(\ -\Delta\ +\ V(x)\ \right)\Phi\ +\ g|\Phi|^2\Phi,
\ \ x\in\R^3
\label{NLS-GPa}\end{align}
We assume that $-\Delta +V$ has \underline{multiple} independent localized eigenstates.
For simplicity we assume two distinct eigenvalues.

As noted, the linear time-evolution ($g=0$), for $t\gg1$, settles  down to a quasi-periodic state, a linear superposition of time-periodic and spatially localized solutions, corresponding to the independent eigenstates.
Note also, that for $g\ne0$ there may be multiple co-existing branches of nonlinear defect states; see, for example, the discussion of bound states for the case where $V$ is a double-well potential,  discussed in section \ref{bound-states}.\medskip

\nit {\bf Question:}  What is the long term ($t\uparrow\infty$) behavior of the NLS / GP ($g\ne0$), \eqref{NLS-GPa}  for initial data of small norm?
\medskip

\nit Our results show that under reasonable conditions (which have been explored experimentally \cite{MLS:05,SW:05}) we have:
\begin{enumerate}
\item {\it Ground state selection\ \cite{SW:04,SW:05,GW:08,GW:11}:}\ The generic large time behavior of the initial value problem is \underline{periodic} and is, in particular, a nonlinear ground state of the system. See also \cite{Buslaev-Perelman:93,Buslaev-Perelman:95,TY:02a,TY:02b,Buslaev-Sulem:03}.
\item {\it Energy equipartition\ \cite{GW:11}:}\ For initial data conditions whose nonlinear ground state and  excited state components are equal in $L^2$, the solution approaches a new nonlinear ground state, whose $L^2$ norm has gained an amount equal one-half that of the initial excited state energy. The other one-half of the excited state energy is radiated away.
\end{enumerate}

To state our results precisely requires some mathematical setup. For simplicity we assume that the linear operator $-\Delta+V$ has
the following properties:
\begin{enumerate}
\item[(V1)]
$V$ is real-valued and decays sufficiently rapidly, {\it e.g.} exponentially,  as $|x|$ tends to infinity.\\
\item[(V2)] The linear operator $-\Delta+V$ has two eigenvalues $E_{0\star}<E_{1\star}<0$.
 $E_{0\star}$ is the lowest eigenvalue with
ground state $\psi_{0\star}>0$, the eigenvalue $E_{1\star}$ is possibly degenerate
with multiplicity $N\ge1$ and corresponding eigenvectors
$\xi_{1\star},\xi_{2\star},\cdot\cdot\cdot,\xi_{N\star}.$
\item[(V3)] Resonant coupling assumption
 \begin{equation} \omega_\star\ \equiv\ 2E_{1\star}-E_{0\star}>0,\ \label{2e1me0}.
\end{equation}
\end{enumerate}

We remark on (V3). An important role is played by the mechanism resonant coupling of discrete and continuum modes and energy transfer from localized states to dispersive radiation. In sections \ref{simple-1} and \ref{simple-2} we introduced and analyzed a simple model of this phenomenon. This model was linear and the resonance was due to a  discrete eigenvalue, $\omega$, embedded in the continuous spectrum. For NLS / GP, \eqref{NLS-GPa} resonant coupling of discrete and continuum modes arises due to higher time-harmonic generation by nonlinearity. The assumption (V3) is made to ensure that this coupling occurs at second order in (small) energy of the solution. However coupling at arbitrary higher order can be studied using the normal form ideas developed in \cite{Gang:07,Cuccagna:08,Bambusi-Cuccagna:11,Cuccagna:12}. \medskip

\subsection{Linearization of NLS/GP about the ground state}

Recall Theorem \ref{RWthm} on the family of nonlinear bound states, $E\mapsto \psi_E$, bifurcating from the ground state. We now consider the linearized  NLS / GP, \eqref{NLS-GPa}, time-dynamics about  this family of solutions. Let
\[\Phi(x,t)=e^{-iE t}\left(\ \psi_E\ + u +\ iv\ \right),\]
 where $u$ and $v$ are real and imaginary parts of the perturbation.  Then the linearized perturbation equation about $\psi_E$, is
\begin{equation}
\frac{\D}{\D t}\left(\begin{array}{ll} u\\ v \end{array} \right)\ =\
 L(E)\ \left(\begin{array}{ll} u\\ v \end{array} \right)\ =\ JH(E)\ \left(\begin{array}{ll} u\\ v \end{array} \right),
 \label{linearized}
 \end{equation}
 where
\begin{equation}\label{eq:opera}
L(E) =
\left(\begin{array}{lll}0&L_{-}(E)\\
 -L_{+}(E)&0 \end{array} \right)
 = \left(\begin{array}{lll}0&1\\
 -1&0 \end{array} \right)\ \left(\begin{array}{lll}L_+(E)&0\\
0& L_{-}(E) \end{array} \right)\ \equiv\ JH(E)\ .
 \end{equation}

The operators $L_+$ and $L_-$ are given by:
\begin{align}
 L_{-}(E)&\ =\ -\Delta-E+V+g(\psi_E)^{2}\nn\\L_{+}(E)&\ =-\Delta-E+V+3g(\psi_{E})^{2}\label{Lpm}
 \end{align}

\nit The following result on the linearized matrix-operator, $L(E)$, proved by standard perturbation theory \cite{RS4}, is given in ~\cite{GW:08} (Propositions 4.1 and 5.1):
\begin{lem}\label{LM:NearLinear}
Assume (V1), (V2) and (V3). Let $L(E)$
denote the linearized operator about a state on the branch of bifurcating bound states, given by Theorem \ref{RWthm}:
\begin{equation}
\psi_E(x)=\rho(E) \left(\ \psi_{0\star}(x)\ +\ \mathcal{O}\left(\ \rho(E)^{2}\right)\  \right),\ \ \textrm{where}\ \
 \rho(E)\equiv\left|E_{0\star}-E\right|^{1\over2}\ \left(\ |g|\ \int\psi_{0\star}^{4}\right)^{-\frac{1}{2}}
\label{psiEexpand1}\end{equation}
For $E=E_{0\star}$, the matrix operator $L(E_{0\star})$ has complex conjugate eigenvalues $\pm i\beta_\star=\pm i(E_{1\star}-E_{0\star})$, each of
multiplicity $N$. For $|E_{0\star}-E|$ and small these perturb to
(possibly degenerate) eigenvalues $\pm i\beta_1(E),\dots,$ $\pm
i\beta_N(E)$ with corresponding neutral  modes (eigenstates)
$$\left(
\begin{array}{lll}
\xi_{1}\\
\pm i\eta_{1}
\end{array}
\right),\ \left(
\begin{array}{lll}
\xi_{2}\\
\pm i\eta_{2}
\end{array}
\right),\ \cdot\cdot\cdot, \left(
\begin{array}{lll}
\xi_{N}\\
\pm i\eta_{N}
\end{array}
\right)$$
\begin{equation}\label{eq:Orthogonality}
\textrm{satisfying}\ \ \langle \xi_{m},\eta_{n}\rangle =\delta_{m,n},\ \langle \xi_{m},\psi_{E}\rangle=\langle \eta_{m},\partial_{E}\psi_{E}\rangle=0\ .
\end{equation}
Moreover,
\begin{equation}\label{eq:GoToNear}
0\not=\displaystyle\lim_{E\rightarrow
E_{0\star}}\xi_{n}=\lim_{E\to E_{0\star}}\eta_{n}\in
{\rm span}\{\xi_{1\star},\ \xi_{2\star},\dots,\ \xi_{N\star}\}\
\end{equation}
 in Sobolev  $H^k$ spaces for any $k>0$.

\nit Furthermore, we note that for $|E_{0\star}-E|$ sufficiently small
\begin{equation}
2\beta_n(E)+E\approx 2(E_{1\star}-E_{0\star})+E_{0\star}=2E_{1\star}-E_{0\star}>0,\
n=1,2,\cdot\cdot\cdot,N,
\label{RES}\end{equation}
\end{lem}

\begin{remark}
Equation \eqref{RES} (see also (V3), \eqref{2e1me0}) ensures coupling of discrete to continuous spectrum at second order in $|E-E_{0\star}|$.
\end{remark}

\subsection{Ground state selection and energy equipartition}

In this section we give a detailed description of the long term evolution.\medskip

\begin{theo}[Ground State Selection]\label{THM:MassTransfer}

Consider NLS / GP, \eqref{NLS-GPa}, with linear potential satisfying (V1), (V2) and (V3), and cubic nonlinearity ($\sigma=1$).
 Assume, most notably, that the non-negative (Fermi golden rule) expression for $\Gamma_0(z,z^*)$ in \eqref{FGR} is strictly positive. (See \cite{SW:04,GW:08,GW:11} for detailed statement of all technical assumptions. This expression is always non-negative and generically strictly positive due to the  assumption \eqref{2e1me0}.\ ).
%

\nit Take initial conditions of the form:
\begin{equation}
\psi_{0}(x)=e^{i\gamma_{0}}\left[\ \psi_{E_{0}}+\alpha_{0}\cdot\xi+i\beta_{0}\cdot \eta +R_{0}\ \right]\ ,
\label{nls-gp-data}
\end{equation}
where  $\gamma_{0}$ and $E_0\in\mathcal{I}=(E_{0\star}-\delta_0,E_{0\star})$ are real constants,   $\alpha_{0}$ and $\beta_{0}$ are real $1\times N$ vectors, and $R_0:\mathbb{R}^{3}\rightarrow \mathbb{C}$,  are such that
\begin{align}
& |E_{0}-E_{0\star}| \ll1,\quad
\frac{ |\alpha_{0}|^{2}+|\beta_{0}|^{2}}{ \|\psi_{E_{0}}\|^2_{2}}\ \ll1,\quad
\|\langle x\rangle^{4}R_{0}\|_{H^2}\lesssim |\alpha_{0}|^{2}+|\beta_{0}|^{2}
\label{data-constraints}\end{align}

 Then there exist smooth functions
  $E(t):\mathbb{R}^{+}\rightarrow
\mathcal{I}$,\  $\gamma(t): \mathbb{R}^{+}\rightarrow
\mathbb{R}$, \  $z(t):\mathbb{R}^{+}\rightarrow \mathbb{C}^{N}$ and $  R(x,t):\mathbb{R}^{3}\times\mathbb{R}^{+}\rightarrow
\mathbb{C}$
 such that the solution of NLS / GP evolves in the form:
\begin{align}\label{Decom}
\psi(x,t)&\ =\ e^{-i\int_{0}^{t}E(s)ds}e^{i\gamma(t)}\nn\\
&\  \times\left[  \psi_{E(t)}\ +\ a_{1}(z,\bar{z})\
\D_E\psi_{E}\ +\ ia_{2}(z,\bar{z})\ \psi_{E} + (Re\ \tilde{z}[z,\bar{z}])\cdot\xi + i\ (Im\tilde{z}[z,\bar{z}])\cdot\eta + R \right],\end{align}
where  $\lim_{t\rightarrow \infty} E(t)=E_\infty,$
for some $E_{\infty}\in \mathcal{I}$.\\
Here, $a_{1}(z,\bar{z}),\ a_{2}(z,\bar{z}): \mathbb{C}^{N}\times\mathbb{C}^{N}\rightarrow \mathbb{R}$ and $\tilde{z}-z: \mathbb{C}^{N}\times\mathbb{C}^{N}\rightarrow \mathbb{C}^{N}$
 are polynomials of $z$ and $\bar{z}$, beginning with terms of order $|z|^{2}$.

\begin{enumerate}
\item[(A)]  The dynamics of mass/energy transfer is captured by the following reduced dynamical system for the  key modulating parameters, $E(t)$ and $z(t)$:
\begin{equation}\label{eq:IncreaseLambda}
\frac{d}{dt}\ \left\|\psi_{E(t)}\right\|_2^2 =z^* \Gamma_0(z,\bar{z})\ z +\ \mathcal{S}_E(t),
\end{equation}
\begin{equation}\label{eq:DecayZ}
\frac{d}{dt}\ |z(t)|^2  = -2z^* \Gamma_0(z,\bar{z})\ z+ \mathcal{S}_{z}(t)\ .
\end{equation}
Here, $\Gamma_0$ is a Fermi golden rule (damping) matrix  given by
\footnote{$\Gamma_0(z,z^*)$ assumed to be strictly positive, is  non-negative by the following:
\begin{align}
&\Im\ [-\Delta+V-\omega_\star-i0]^{-1} = \frac{1}{2i}\ \lim_{\delta\downarrow0}\ \left(\
  [-\Delta+V-\omega_\star-i\delta]^{-1}
   -  [-\Delta+V-\omega_\star+i\delta]^{-1}
   \ \right),\nn\\
&=\ \ \pi\ \delta\left(-\Delta+V-\omega_\star\right),\ \ {\rm where}\ \omega_\star\in\sigma_{cont}(-\Delta+V).\nn
\end{align} Note: $\Im\ [-\Delta+V-\omega_\star-i0]^{-1}$ projects onto the generalized mode at energy $\omega_\star\in\sigma_{cont}(-\Delta+V).$
}
\begin{equation}
\Gamma_0(z,z^*) \equiv g^2\ \Im \left\langle [-\Delta+V-\omega_\star-i0]^{-1}\ \psi_{E}\ (z\cdot\xi)^{2},\
 \psi_{E} \ (z\cdot\xi)^{2}\right\rangle\ \ge\ c^2\ |z|^4
\label{FGR}\end{equation}
 and
$S_E(t),\ S_z(t)\ \lesssim\ (1+t)^{-\rho}$, with
$\int_0^\infty\ S_E(\tau)d\tau$\ and\ $\int_{0}^{\infty} \mathcal{S}_z(\tau) d\tau\ =\  o(|z_0|)^2.$
 \item[(B)] {\bf Estimates on $z(t)$ and the correction $\vec{R}(t)$:}
 For all $t\geq 0$, we have $\|\vec{R}(t)\|_{H^2}\leq \epsilon_{\infty}$. Moreover,  the following decay estimates hold:
 \begin{align}
&\left\|(1+x^{2})^{-\nu}\vec{R}(t) \right\|_{2}\ \leq\ C\ (\|\langle x\rangle^{4}\psi_0\|_{H^2})\ (1+t)^{-1},
\label{Rdecay}\\
&|z(t)|\ \leq\ C\ (\|\langle x\rangle^{4}\psi_0\|_{H^2})\ (1+t)^{  -\frac{1}{2}  }
\label{z-decay}
\end{align} 
\end{enumerate}
\end{theo}

\begin{theo}[ Mass / Energy equipartition]\label{mass-equipartition}\ Consider NLS / GP, \eqref{NLS-GPa},  under the technical hypotheses of Theorem \ref{THM:MassTransfer} and assumptions on the initial data \eqref{data-constraints},
 where $|\alpha_0|^2 +|\beta_0|^2$ is a measure of the neutral modes'  perturbation of the ground state. Recall that $E_0$ is the energy of the initial nonlinear ground state and $E_\infty$ is the asymptotic energy, guaranteed by Theorem \ref{THM:MassTransfer}. Then, in the limit as $t\to\infty$, one half of the  neutral modes' mass contributes to forming a more massive asymptotic ground state and one half is radiated away as dispersive waves:
\begin{equation}\label{eq:Mass}
\|\psi_{E_{\infty}}\|_{2}^{2}=
\|\psi_{E_{0}}\|_{2}^{2}+\frac{1}{2}\left[\ |\alpha_{0}|^{2}+|\beta_{0}|^{2}\ \right]
+o\left(\ |\alpha_{0}|^{2}+|\beta_{0}|^{2}\ \right).
\end{equation}
\end{theo}

\begin{figure}[ht]
\begin{center}
\includegraphics[height=3in]{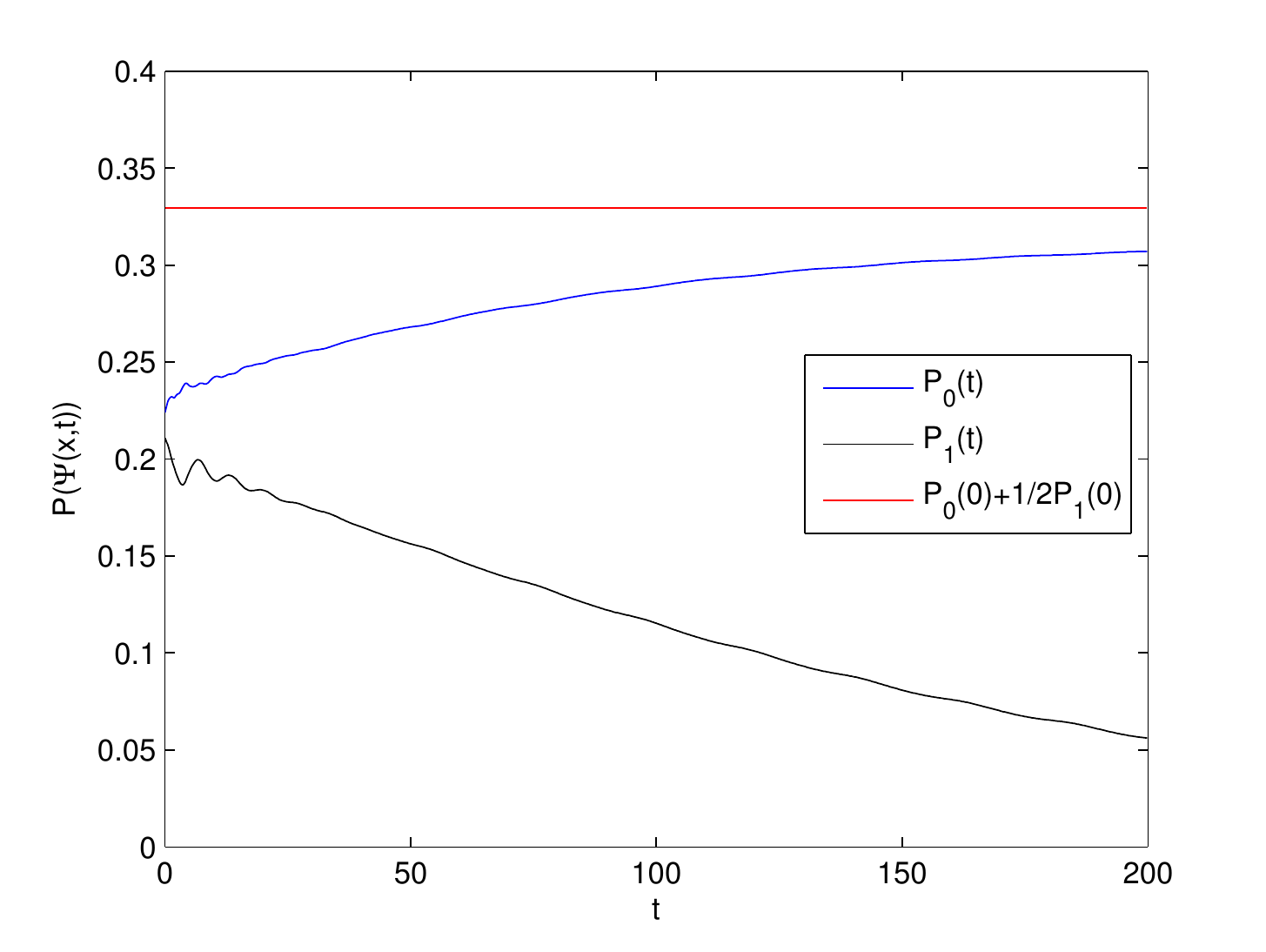}
\end{center}
\caption{{\footnotesize{Numerical verification ground state selection and asymptotic energy equipartition. \ Computations performed by E. Shlizerman\ .}}}
\label{en-equi-fig}
\end{figure}

Figure \ref{en-equi-fig} illustrates the phenomena of ground state selection and mass / energy equipartition in NLS / GP.  These simulations are for the case of double-well potentials (see section \ref{double-well}). Asymptotic energy equipartition  has been verified for the case when the stable ground state is symmetric ($\cN<\cN_{\rm cr}$) and the case when the stable ground states is asymmetric ($\cN>\cN_{cr}$).

A result on equipartition of energy holds as well, for NLS-GP with a general power nonlinearity, \eqref{NLS-GP}, under more restrictive hypotheses \cite{GW:11}. Finally, we remark that it would be of interest to establish detailed results on energy-transfer in systems with multiple bound states for  subcritical nonlinearities.  In this setting the current perturbative treatment of small amplitude dispersive waves does not apply. A step in this direction is in the work
 \cite{Kirr-Zarnescu:07,Kirr-Zarnescu:09,Kirr-Mizrak:09}, where asymptotic stability for systems with a single family of nonlinear bound states is treated by a novel  time-dependent linearization procedure.

\section{A nonlinear toy model of nonlinearity-induced energy transfer }

In this section we explain the idea behind the results on ground state selection and energy equipartition. In the style of sections \ref{simple-1} and \ref{simple-2} we introduce a toy minimal model, here nonlinear, which captures the essential mechanisms and is, by comparison, simpler to analyze.

Our model is for the interaction of three amplitudes:
\begin{enumerate}
\item a ``ground state'' complex amplitude, $\alpha_0(t)$ with associated frequency $E_{0\star}$
\item an ``excited state" complex amplitude, $\alpha_1(t)$ with associated frequency $E_{1\star}$, and
\item a ``continuum wave-field'' complex amplitude, $R(x,t)$, with spectrum of frequencies given by the positive real-line $[0,\infty)$.
\end{enumerate}
As with NLS / GP, our model has cubic nonlinearity and the assumption
\begin{equation}
\omega_\star\equiv 2E_{1\star}-E_{0\star}>0
\label{omega-res}
\end{equation}
is assumed to assure coupling of discrete and continuum modes at second order in the solution amplitude.
We also introduce a  function, $\chi(x)$, sufficiently rapidly decaying at spatial infinity. (Recall that in  section \ref{simple-1} we took $\chi(x)=\delta(x)$.)

Our nonlinear model is the following:
\begin{align}
&i\D_t\alpha_0(t)\ -\ \Omega_0\ \alpha_0=\ g\ \left\langle\chi,\overline{R(t)}\right\rangle\ \alpha_1^2(t)\label{nl-toy0}\\
&i\D_t\alpha_1(t)\ -\ \Omega_1\ \alpha_1 =\ 2g\ \left\langle\chi,R(t)\right\rangle\   \alpha_0(t)\ \overline{\alpha_1(t)}\label{nl-toy1}\\
&i\D_t\ R(x,t)\ =\ -\Delta\ R(x,t)\ +\ g\ \chi(x)\ \alpha_1^2(t)\ \overline{\alpha_0(t)}
\label{nl-toyR}\end{align}

A first observation is that the system \eqref{nl-toy0}-\eqref{nl-toyR} is Hamiltonian and has the time-invariant quantity
\begin{equation}
\frac{d}{dt}\ \left(\ |\alpha_0(t)|^2\ +\ |\alpha_1(t)|^2\ +\ \int_{\R^d}\ |R(x,t)|^2\ dx\ \right)\ =\ 0\nn
\end{equation}

To solve \eqref{nl-toy0}-\eqref{nl-toyR}, we first separate fast and slow scales by introducing slowly varying amplitudes, $A_0$ and $A_1$:
\begin{equation}
\alpha_0(t)\ =\ e^{-i \Omega_0 t}\ A_0(t),\ \ \alpha_1(t)\ =\ e^{-i \Omega_1t}\ A_1(t)\nn\end{equation}
Then, the system for $A_0, A_1$ and $R$ becomes:
\begin{align}
&i\D_t A_0(t)\ =\ g\ \left\langle\chi,\overline{R(t)}\right\rangle\ e^{-i\omega_\star  t}A_1^2(t)\label{nl-toyA0}\\
&i\D_t A_1(t)\ =\ 2g\ \left\langle\chi,R(t)\right\rangle\  e^{i\omega_\star t}\ A_0(t) \overline{A_1(t)}\label{nl-toyA1}\\
&i\D_t\ R(x,t)\ =\ -\Delta\ R(x,t)\ +\ g\ \chi(x)\ e^{-i\omega_\star t}\ A_1^2(t)\ \overline{A_0(t)}\label{nl-toyRA}
\end{align}

Note that since the spectrum of $-\Delta=[0,\infty)$ and since  $\omega_\star>0$ (by assumption)
the Schr\"odinger wave equation, \eqref{nl-toyRA} is resonantly forced.
We proceed now to solve for $R(x,t)$, keeping its dominant contributions.
 Duhamel's principle and  a regularization, motivated by the local decay estimate \eqref{ld-est} below,   gives:
\begin{align}
&R(t)= e^{i\Delta t}R(0)  -ig\ e^{i\Delta t} \int_0^t\ e^{-i(-\Delta+\omega_\star)s}\ \chi \overline{A_0(s)}\ A_1^2(s)\ ds\nn\\
&=^{\delta\downarrow0} \ e^{i\Delta t}R(0)\  -ig\ e^{i\Delta t}\ \int_0^t \frac{1}{i(-\Delta+\omega_\star-i\delta)}\ \frac{d}{ds}\left[ e^{i(-\Delta-\omega_\star)s}\right]\chi \overline{A_0(s)} A_1^2(s)\ ds
\nn\end{align}
Next we have, by integration by parts,
\begin{align}
&R(x,t)\ \sim\  e^{i\Delta t}R(0)\ -\  ge^{-i\omega_\star t} \frac{1}{-\Delta-\omega_\star-i0}\ \chi\
 \overline{A_0(t)}A_1^2(t)\ \nn\\
&\ \ +\  \overline{A_0(0)}A_1^2(0)\ \frac{e^{i\Delta t}}{-\Delta-\omega_\star-i0}\ \chi\ \nn\\
&\ \ +\ \int_0^t\ \frac{e^{i\Delta (t-s)}}{-\Delta-\omega_\star-i0}\ \chi\ \frac{d}{ds}\left[\overline{A_0(s)} A_1^2(s)\right]\ ds\ ,
\nn\end{align}
which may be written as
\begin{align}
&R(t)\ \sim\ -ge^{-i\omega_\star t}\
\frac{1}{-\Delta-\omega_\star-i0}\ \chi\ \overline{A_0(t)}A_1^2(t)\ +\ \cO\left( t^{-{3\over2}}
\right)\label{R1expand}\end{align}
where the error term is obtained using the  {\it local energy decay estimate}:
 \begin{equation}\label{ld-est}
 \left\|\langle x\rangle^{-\nu}\ \frac{e^{i\Delta t}}{-\Delta-\omega_\star-i0}\ \langle x\rangle^{-\nu}\right\|_{L^2(\R^3)\to L^2(\R^3)}\ =\ {\cO(t^{-{3\over2}}),\ \ t\to+\infty,}
 \end{equation}
 where $\nu$ is sufficiently large and positive.
 Substitution of the leading order terms of $R(t)$ in the expansion \eqref{R1expand} into \eqref{nl-toyA0}-\eqref{nl-toyA1} and use of the distributional identity
 \[ (-\Delta-\omega_\star-i0)^{-1}\ =\ {\rm P.V.}\ \frac{1}{-\Delta-\omega_\star}\ +\ i\pi\ \delta(-\Delta-\omega_\star)\]
yields
the \underline{dissipative} system for the amplitudes $A_0$ and $A_1$:
\begin{align}
\D_t A _0\ &\approx\ \ +\ {\Gamma}\ |A_1|^4\ A_0\ \ ,\ \ \D_t A_1\ \approx\ \ -2\Gamma\ |A_0|^2|A_1|^2 A_1\label{damp-ds}
\end{align}
where  $\Gamma\ \sim\ g^2\ \left|\ \widehat{\chi}(\omega_*)\ \right|^2\ \ge\ 0$
and is generically strictly positive. This is the analogue of the strict positivity discussed in the statement of Theorem \ref{mass-equipartition}. Here, $\widehat{\chi}$ denotes the Fourier transform of $\chi$.

The omitted correction  terms  in \eqref{damp-ds} can be controlled in terms of the quantities
$\| R(t)\|_{L^p}$ ($p>2$ and sufficiently large),\  $\| \langle x\rangle^{-\nu} R(t)\|_{L^2}$ and  $\|\  \langle x\rangle^{-\nu}e^{i\Delta t}(-\Delta-\omega_*-i0)^{-1}\ \langle x\rangle^{-\nu}\|_{L^2\to L^2}$.\medskip

Finally, ground state selection with energy equipartition can be seen through the reduction \eqref{damp-ds}. Indeed, setting  $P_0=|\alpha_0|^2,\ \ P_1=|\alpha_1|^2$ we obtain
\begin{equation}
\frac{dP_0}{dt}\ \sim {\Gamma} P_1^2\ P_0\ ,\qquad
\frac{dP_1}{dt}\ \sim -2{\Gamma} P_1^2\ P_0
 \nn\end{equation}
 It follows that $ 2P_0(t)+P_1(t) \sim 2P_0(0)+P_1(0)$. Sending $t\to\infty$ and using that $P_1(t)\to0$,\ (indeed, $\Gamma>0$), we obtain $P_0(\infty)  = P_0(0) + \frac{1}{2}P_1(0)$.

\section{Concluding remarks}

A fundamental question in the theory of nonlinear waves for  non-integrable PDEs is whether and in what sense
arbitrary finite-energy initial conditions evolve toward the family of available nonlinear bound states
and radiation. This question is often referred to as the {\it Soliton Resolution Conjecture}; see, for example, \cite{Tao:09}. The asymptotic stability / nonlinear scattering results of the type discussed in section \ref{gss-ep}, being
 based on normal forms ideas, spectral theory of linearized operators, dispersive estimates and low-energy (perturbative) scattering methods, give a local picture of the phase space near families of nonlinear bound states.  In contrast, integrable nonlinear systems which  can be mapped to exactly solvable linear motions, allow for a global description of the dynamics \cite{Deift-Its-Zhou:93,Deift-Zhou:03,Deift-Park:11,Cuccagna-Pelinovsky:13}.
  We also note the important line of research by Merle {\it et. al.}  which is based on the monotone evolution properties of appropriately designed local energies (see, for example, in  \cite{Martel-Merle:01,
 Merle-Raphael:04}) and does not rely on dispersive time-decay estimates of the linearized flow.  It would be of great interest to adapt these methods for use in tandem with dynamical systems ideas to situations where
the background linear medium, {\it i.e.} a potential $V(x)$,  gives rise to structures (defect modes) which interact with ``free solitons''.\medskip

With a view toward understanding asymptotic resolution in non-integrable Hamiltonian PDEs, one of the simplest global questions to ask is the following. Let  $V$ denote a smooth, radially-symmetric and uni-modal potential well which decays very rapidly at spatial infinity. Assume further that
 $-\Delta+V$ has a finite number of bound states. Consider NLS / GP with a \underline{repulsive} nonlinear potential ($g=+1$) with radially symmetric initial conditions:
 \begin{align*}
&  i\D_t \Psi =-\Delta \Psi\  +\ V(x)\Psi  +  |\Psi|^{p-1}\Psi,\ \ \ \Psi(x,0)\ =\ \Psi_0(|x|)\ \in\ H^1_{radial}(\R^d)
\end{align*}
It is natural to expect that energy which remains spatially localized must reside the family of nonlinear defect modes, {\it i.e.} bound states, localized within the potential well, which lie on solution branches bifurcating from the linear bound states of $-\Delta +V$ \cite{RW:88}; see section \ref{bound-states}. Indeed, any concentration of energy outside the well should disperse to zero, since for $|x|$ large the equation is a translation invariant NLS equation with repulsive (defocusing) nonlinearity having no bound states. We conjecture that for arbitrary initial data solutions either disperse to zero or approach a nonlinear defect state,
and furthermore that generic initial conditions approach a stable nonlinear defect state.

A step in this direction is the result of Tao \cite{Tao:08} for $1+\frac{4}{d}<p<2^\star$,
in very high spatial dimensions. {\it For spatial dimensions $d\ge 11$ there is a compact attractor. That is, there exists $K$, a  compact subset of $H^1_{radial}(\R^d)$, such that $K$ is invariant under the the NLS / GP flow.
 Moreover, if $u\in C^0_tH^1_x(\R\times\R^d)$  is a global-in-time solution of NLS / GP, then there exists $u_+\in H^1(\R^d_{radial})$ such that
 $dist_{H^1}\left( \Psi(t)-e^{i\Delta t}u_+ , K\ \right)\to0,\ \ \ {\rm as}\ \ \ t\to+\infty$.}

\nit It is natural to conjecture that
 $K$ is the set of nonlinear bound (defect) states.
 \medskip

 Finally we mention that the detailed dynamical picture of energy transfer from discrete to radiation modes has connections with important questions in applied physics. Note that the Fermi golden rule damping matrix, $\Gamma_0(z,z)$, appearing in Theorems \ref{THM:MassTransfer} and \ref{mass-equipartition}, which controls the rate of decay of excited states (neutral modes) is controlled by the density of states of the linearized operator near the resonant frequency $\omega_\star$; see \eqref{2e1me0}.
 \smallskip

\nit {\it Control problem:\ Can one design the potential, $V(x)$, in NLS / GP  in order to inhibit / enhance energy transfer and relaxation (decay) to the system's asymptotic state?}

This relates to the problem of  controlling the spontaneous emission rate of atoms by modifying the background environment of the atom (for us, the background linear potential), thereby controlling the density of states \cite{Cohen-T:92}.  A problem of this type was investigated analytically and computationally for a closely related parametrically forced linear Schr\"odinger equation \cite{OW:11}. Such a study for NLS / GP  could inform the design of experiments, such as those reported on in \cite{MLS:05}.

\bibliographystyle{plain}
\bibliography{fiads}
\end{document}